\documentclass[showpacs,preprintnumbers,twocolumn,eqsecnum,superscriptaddress,aps,nofootinbib]{revtex4}
\usepackage{amssymb}
\usepackage[centertags]{amsmath}
\usepackage{txfonts}
\usepackage{epsfig}
\usepackage{bm}
\usepackage{color}
\usepackage{graphicx,graphics}
\usepackage{multirow}
\usepackage{float}
\usepackage[pdfstartview=FitH]{hyperref}
\hypersetup{colorlinks=true,citecolor=blue,linkcolor=blue,filecolor=black,urlcolor=blue}
\allowdisplaybreaks[2]

\usepackage{slashed}
\usepackage{ulem}

\usepackage{booktabs}

\begin{document}
\title{Three Dimensional Imaging of the Nucleon and Semi-Inclusive High Energy Reactions}

\author{Kai-bao Chen}
\affiliation{School of Physics \& Key Laboratory of Particle Physics and Particle Irradiation (MOE), Shandong University, Jinan, Shandong 250100, China}

\author{Shu-yi Wei}
\affiliation{School of Physics \& Key Laboratory of Particle Physics and Particle Irradiation (MOE), Shandong University, Jinan, Shandong 250100, China}

\author{Zuo-tang Liang}
\affiliation{School of Physics \& Key Laboratory of Particle Physics and Particle Irradiation (MOE), Shandong University, Jinan, Shandong 250100, China}

\begin{abstract}
We present a short overview on the studies of transverse momentum dependent parton distribution functions of the nucleon.
The aim of such studies is to provide a three dimensional imagining of the nucleon and a comprehensive description of semi-inclusive high energy reactions. 
By comparing with the theoretical framework that we have for the inclusive deep inelastic lepton-nucleon scattering and the one-dimensional imaging of the nucleon, 
we summarize what we need to do in order to construct such a comprehensive theoretical framework for semi-inclusive processes in terms of three dimensional gauge invariant parton distributions. 
After that, we present an overview of what we have already achieved with emphasize on the theoretical framework for semi-inclusive reactions 
in leading order perturbative QCD but with leading and higher twist contributions. 
We summarize in particular the results for the differential cross section and the azimuthal spin asymmetries in terms 
of the gauge invariant transverse momentum dependent parton distribution functions.
We also briefly summarize the available experimental results on semi-inclusive reactions and parameterizations of transverse momentum dependent parton distributions 
extracted from them and make an outlook for the future studies.
\end{abstract}



\maketitle

\section{Introduction}
With the deeply going of the study of the nucleon structure, three dimensional imaging has become the very frontier and a hot topic in recent years.   
It is commonly recognized that the three dimensional imaging contains much more abundant physics  on the nucleon structure and the properties 
of quantum chromodynamics (QCD). 
The study was initially triggered by the experimental finding of striking single-spin asymmetries (SSA) 
in inclusive hadron production in hadron-hadron collisions 
with transversely polarized hadron~\cite{SSA}.
Gradually it grows into a field aiming at a comprehensive three dimensional description of the nucleon structure 
including spin and transverse momentum dependences.

The one dimensional imaging of the nucleon is provided by the Parton Distribution Functions (PDFs) such as the number densities, $q(x)$, 
the helicity distributions, $\Delta q(x)$, and the transversities, $\delta q(x)$, for quarks of different flavors in the nucleon. 
These one dimensional PDFs can be studied in inclusive high energy reactions and are necessary for the description of such inclusive processes.  
In the three dimensional case, i.e. where the parton transverse momentum is also considered, not only the direct extensions of these distribution functions 
to include transverse momentum dependences are involved, but also many other correlation functions that describe in particular the correlations between 
the transverse momenta and spins such as the Sivers function, the Boer-Mulders function, the pretzelocity etc. exist. 
They are generally called transverse momentum dependent (TMD) PDFs. 
Moreover, higher twist effects become also important and need to be considered consistently. 
The content of the studies is therefore much more abundant and more interesting.
These TMD PDFs can be studied in semi-inclusive reactions and are necessary for the description of such processes. 

The study on the three dimensional imaging of the nucleon is in a rapid developing phase and it is not so easy to make 
a comprehensive overview of all different aspects of the studies. 
Here, we choose to arrange the review in the following way: 
First we will make a brief review of what we did in one dimensional case with inclusive deep inelastic lepton-nucleon scattering (DIS). 
In this way, we hope that we can find out the main line of what we need to do in three dimensional case. 
After that we will try to summarize the progresses already achieved along this line and what we need to do next. 
Such a brief review of the one dimensional case will be presented in Sec. 2. 
In Sec. 3, we will make a short summary of TMDs defined via quark-quark correlator.
In Sec. 4, we will present a brief overview of what we have for constructing the theoretical framework of semi-inclusive processes. 
In Sec. 5, we will make a short summary of the available experimental results and TMD parameterizations extracted from them. 
Finally we will make a short summary of this review in Sec. 6. 

This overview article is an extended version of a plenary talk at the 21st international symposium on spin physics (Spin2014)~\cite{Liang:2015nia}. 
As can be imagined that the simplest and basic picture is what we have at the leading order in perturbative QCD (pQCD) and at the leading twist. 
Hence, there are also two major directions in theoretical developments towards a comprehensive description of the semi-inclusive processes. 
One is to take higher order pQCD into account, and the second is to  consider higher twist contributions. 
These contributions are important not only for higher accuracy but also for consistency.  
The major progresses that have been made in recent years are also in these two directions separately, 
i.e. either at the leading twist but leading and higher order in pQCD or leading order in pQCD but leading and higher twists. 
The talk~\cite{Liang:2015nia} was mainly concentrated on the second direction. 
For higher order pQCD contributions where evolutions of PDFs are involved, an overview talk was also presented by Daniel Boer in the same conference~\cite{Boer:2015ala}. 
There are also many other reviews and monographs (e.g. \cite{Collins:2013zsa,Collins:1989gx,Collins:2011zzd}). 
The study for higher order in pQCD and higher twists seems to be rather difficult and even the factorization properties are unclear~\cite{nonfact}.  
In this article, we follow the same line as in the talk~\cite{Liang:2015nia} but briefly summarize the progresses in the studies on QCD evolutions
and refer the interested readers to those reviews.

\section{Inclusive DIS \& the One Dimensional Imagining of the Nucleon}

Our studies on the structure of a fast moving nucleon started with inclusive DIS such as $e^-+N\to e^-+X$. 
We recall that, under one photon exchange approximation, the differential cross section is given by the Lorentz contraction of 
the well-known leptonic tensor $L^{\mu\nu}(l,l',\lambda_l)$ and 
the hadronic tensor $W_{\mu\nu}(q,p,S)$, i.e.,
\begin{equation}
d\sigma=\frac{2\alpha_{\rm em}^2}{sQ^4}L^{\mu\nu}(l,l',\lambda_l)W_{\mu\nu}(q,p,S)\frac{d^3l'}{2E_{l'}}.
\end{equation}
The leptonic tensor is calculable and is given by,
\begin{equation}
L_{\mu\nu}(l,l',\lambda_l) = 2 (l_\mu l'_\nu + l_\nu l'_\mu - g_{\mu\nu} l \cdot l') + i2\lambda_l\epsilon_{\mu\nu\rho\sigma}l^\rho q^\sigma.
\end{equation}
Information on the structure of the nucleon is contained in the hadronic tensor defined as,
\begin{align}
W_{\mu\nu}(q,p,S)=\frac{1}{2\pi} & \sum_X\langle p,S\left|j_\mu(0)\right|X\rangle\langle X\left|j_\nu(0)\right|p,S\rangle \nonumber\\
&\times (2\pi)^4\delta^4(p+q-p_X).
\end{align}
Here, $l$ and $p$ denote the 4-momenta of the lepton and the nucleon respectively, those with prime are for the final states;
$\lambda$ stands for the helicity,  and $S$ for the polarization vector of the nucleon. 
We use the light-cone coordinate and 
define the light-cone unit vectors as $\bar n = (1,0,\vec 0_\perp)$, $n =(0,1,\vec 0_\perp)$, $n_\perp = (0,0,\vec n_\perp)$, 
so that a general four-vector can be decomposed as $A^\mu = A^+ \bar n^\mu + A^- n^\mu + A_\perp^\mu$, 
with $A^{\pm} = (A^0 \pm A^{3})/\sqrt{2}$, and $A_\perp = (0,0,\vec A_\perp)$. 
We work in the center of mass frame of the $\gamma^*N$ and choose the nucleon's momentum as $z$-direction 
so that $p$ and $S$ are decomposed as,
\begin{align}
& p^\mu = p^+ \bar n^\mu + \frac{M^2}{2p^+} n^\mu,\\
& S^\mu = \lambda \frac{p^+}{M}\bar n^\mu + S_T^\mu - \lambda \frac{M}{2p^+} n^\mu.
\end{align}
The Bjorken variable is defined as $x_B=Q^2/2p\cdot q$,  $q=-x_Bp^+\bar n + nQ^2/(2x_Bp^+)$; 
and we also define $y=p\cdot q/p\cdot l$.

The theoretical framework for inclusive DIS has been constructed in the following steps. 
First, we studied the kinematics and obtained the general form of the hadronic tensor by applying the 
basic constraints from the general symmetry requirements such as Lorentz covariance, 
gauge invariance, parity conservation and Hermiticity, e.g.,
\begin{align}
&q^\mu W_{\mu\nu}(q,p,S) = 0,\\
&W_{\mu\nu}(\tilde q, \tilde p, -\tilde S) = W^{\mu\nu}(q,p,S),\\
&W_{\mu\nu}^*(q,p,S) = W_{\nu\mu}(q,p,S),
\end{align}
where $\tilde A$ denotes the results of $A$ after space reflection, i.e., $\tilde A^\mu = A_\mu$.
The general form of the hadronic tensor is given by the sum of a symmetric part and an antisymmetric part, 
\begin{equation}
W_{\mu\nu}(q,p,S) = W_{\mu\nu}^{(S)}(q,p) + iW_{\mu\nu}^{(A)}(q,p,S),
\end{equation}
where $W_{\mu\nu}^{(S)}(q,p)$ and $W_{\mu\nu}^{(A)}(q,p,S)$ are given by,
\begin{align}
W_{\mu\nu}^{(S)}&(q,p)=2(-g_{\mu\nu}+\frac{q_\mu q_\nu}{q^2})F_1(x,Q^2) \nonumber\\
& +\frac{1}{xQ^2}(q_\mu+2xp_\mu)(q_\nu+2xp_\nu)F_2(x,Q^2),\\
W_{\mu\nu}^{(A)}&(q,p,S)= \frac{2M\varepsilon_{\mu\nu\rho\sigma}q^\rho}{p\cdot q}  \nonumber\\
&\times\Bigl\{S^\sigma g_1(x,Q^2)+ (S^\sigma-\frac{S\cdot q}{p\cdot q}p^\sigma)g_2(x,Q^2)\Bigr\},
\end{align}
respectively. 
We found out that the hadronic tensor is determined by four independent structure functions $F_1$, $F_2$, $g_1$ and $g_2$, 
where the first two describe the unpolarized case and the latter two are needed for polarized cases.

Our knowledge of one dimensional imaging of the nucleon starts with the ``intuitive parton model" that is very nicely formulated e.g. in \cite{Fey72}. 
Here, it was argued that, in a fast moving frame, because of time dilation, quantum fluctuations such as vacuum polarizations 
can exist quite long. In the infinite momentum frame, such fluctuations exist forever. In this case, a fast moving nucleon can be 
viewed as a beam of free ``partons". 
The probability of the scattering of an electron with a nucleon is taken as the incoherent sum of  that of the scattering with each individual parton, 
more precisely, a convolution of the number density of the parton in the nucleon with the probability of the scattering with the parton, i.e., 
\begin{equation}
|{\mathcal M}(eN\to eX)|^2=\sum_q\int dx f_q(x) |\hat{\cal M}(eq\to eq)|^2 \label{eq:intuitive},
\end{equation}
where $f_q(x)$ is the number density of parton of flavor $q$ in the nucleon. 
In this way, we obtained the famous results~\cite{Fey72}, 
\begin{align}
&F_2(x,Q^2)=2xF_1(x,Q^2)=\sum_q e_q^2xf_q(x), \\
&g_1(x,Q^2)=\sum_q e_q^2\Delta f_q(x),\\ 
&g_1(x,Q^2)+g_2(x,Q^2)=\sum_q e_q^2x\delta f_q(x).
\end{align}
Here, we would like to point out that, with this intuitive parton model, we are doing nothing else but the impulse approximation that we often use
in describing a collision process where we do the following approximations,
\begin{itemize}
\item during the interaction of the electron with the parton, interactions between the partons are neglected;
\item the electron interacts only with one single parton each time;
\item the scatterings of the electron with different partons are added incoherently.
\end{itemize}

Although the physical picture of the intuitive model is very clear and the model is elegant and practical, 
we are not satisfied with the formulation because it is partly qualitative or semi-classical hence 
it is not easy to control the accuracy.  
A proper formulation should be based on quantum filed theory (QFT) and is obtained by starting with the Feynman diagram Fig.~1(a). 
Here, from this diagram, we obtain immediately that, 
\begin{equation}
W_{\mu\nu}^{(0)}(q,p,S)=\frac{1}{2\pi}
\int\frac{d^4k}{(2\pi)^4} 
{\rm Tr}[\hat H_{\mu\nu}^{(0)}(k,q)\hat \phi^{(0)}(k,p,S)],\label{eq:Wincl0}
\end{equation}
where $k$ is the 4-momentum of the parton.
\begin{align}
\hat H_{\mu\nu}^{(0)}(q,k)=\gamma_\mu(\slash{\hspace{-5pt}k}+\slash{\hspace{-5pt}q})\gamma_\nu (2\pi)\delta_+((k+q)^2),  \label{eq:h0}
\end{align}
is a calculable hard part. The matrix element 
\begin{equation}
\hat\phi^{(0)}(k,p,S) =  \int d^4ze^{ik\cdot z} \langle p,S|\bar\psi(0)\psi(z)|p,S\rangle, \label{eq:phi0}
\end{equation}
is known as the quark-quark correlator describing the structure of the nucleon.  
By taking the collinear approximation, i.e. taking $k\approx xp$, and neglecting 
the power suppressed contributions i.e. the $o(M/Q)$ terms, we obtain
\begin{align}
W_{\mu\nu}^{(0)}(q,p) \approx & \Big[ (-g_{\mu\nu}+\frac{q_\mu q_\nu}{q^2})  +\frac{(q+2xp)_\mu(q+2xp)_\nu}{2xp\cdot q} \Big] f_q(x).
\end{align}
This is exactly the same result as that obtained from Eq. (\ref{eq:intuitive}) based on the intuitive parton model.
At the same time, we obtain the QFT operator expression of $f_q(x)$ defined via the quark-quark correlator given 
by Eq.~(\ref{eq:phi0}) as,  
\begin{eqnarray}
f_q(x)=\int \frac{dz^-}{2\pi}e^{ixp^+z^-}\langle p|\bar\psi(0)\frac{\gamma^+}{2}\psi(z)|p\rangle. 
\end{eqnarray}
By inserting the expanded expression of the field operator $\psi(z)$ in terms of the plan wave and the creation and/or annihilation operators, 
we see clearly that $f_q(x)$ is indeed the number density of parton in the nucleon. 
However, from this expression, we see also immediately a severe problem, i.e. this expression is {\it not} (local) gauge invariant! 
We understand that the physical quantity has to be gauge invariant and therefore have to find a solution for this.

\begin{figure}[!ht]
\includegraphics[width=0.45\textwidth]{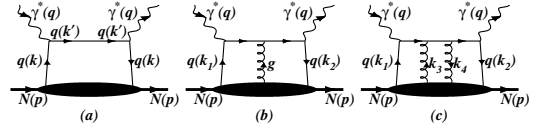}
\caption{Examples of the Feynman diagram series with multiple gluon scattering considered 
 for $\gamma^*+N\to q+X$ with (a) $j=0$, (b) $j=1$ and (c) $j=2$ gluons exchanged.}
\end{figure}

The gauge invariant formulation is obtained by taking into account the multiple gluon scattering shown by the diagram series in Fig.1(a-c). 
This is clear since (local) gauge invariance implies the existence of the gauge interaction that needs to be taken into account. 
In this way, we obtain, 
\begin{eqnarray}
&& W_{\mu\nu}(q,p,S) = \sum_{j=0}^\infty W_{\mu\nu}^{(j)}(q,p,S), 
\end{eqnarray}
where $W_{\mu\nu}^{(j)}(q,p,S)$ represents the contribution from the diagram with exchange of $j$-gluon(s). 
They are all expressed as a trace of a calculable hard part and a matrix element depending on the structure of the nucleon. 
E.g., corresponding to Fig. 1(b), we have $j=1$, and  $W_{\mu\nu}^{(1)}(q,p,S)$ is given by,
\begin{widetext}
\begin{align}
& W_{\mu\nu}^{(1)}(q,p,S) = \sum_{c=L,R} W_{\mu\nu}^{(1,c)}(q,p,S), \\
& W_{\mu\nu}^{(1,c)}(q,p,S)=\frac{1}{2\pi}
\int\frac{d^4k_1}{(2\pi)^4} \frac{d^4k_2}{(2\pi)^4} 
{\rm Tr}[\hat H_{\mu\nu}^{(1,c)}(k_1,k_2,q) \hat \phi^{(1)}_\rho(k_1,k_2,p,S)],\label{eq:Wincl1L}\\
&  \hat \phi^{(1)}_\rho(k_1,k_2,p,S)=\int d^4z d^4y e^{ik_1z+(k_2-k_1)y} \langle p,S|\bar\psi(0)A_\rho(y)\psi(z)|p,S\rangle, \label{eq:phi1}
\end{align}
where $c$ in the superscript represents different cuts (left or right) in the diagram. 
Similarly, corresponding to Fig. 1(c), we have,
\begin{align}
& W_{\mu\nu}^{(2)}(q,p,S) = \sum_{c=L,M,R} W_{\mu\nu}^{(2,c)}(q,p,S), \\
& W_{\mu\nu}^{(2,c)}(q,p,S)=\frac{1}{2\pi}\int\frac{d^4k_1}{(2\pi)^4}\frac{d^4k_2}{(2\pi)^4}\frac{d^4k}{(2\pi)^4} {\rm Tr}[\hat H_{\mu\nu}^{(2,c)\rho\sigma}(k_1,k_2,k,q)
\hat \phi^{(2)}_{\rho\sigma}(k_1,k_2,k,p,S)],\\
& \hat\phi^{(2)}_{\rho\sigma}(k_1,k_2,k,p,S) =
\int d^4yd^4y'd^4ze^{ik_1y+ik(y'-y)+ik_2(z-y')} \langle p,S|\bar\psi(0)gA_\rho(y)gA_\sigma(y')\psi(z)|p,S\rangle.
\end{align}
The matrix element is now a quark-$j$-gluon(s)-quark correlator.
We also immediately see that none of such quark-$j$-gluon(s)-quark correlators is gauge invariant.

To get the gauge invariant form, we need to apply the collinear expansion proposed in Refs.\cite{Ellis:1982wd,Ellis:1982cd,Qiu:1990xxa}, 
which is carried out in the following four steps.

(1) Make Taylor expansions of all hard parts at $k_i = x_ip$, e.g., 
\begin{align}
&\hat H_{\mu\nu}^{(0)}(k,q) =  \hat H_{\mu\nu}^{(0)} (x) + \frac{\partial \hat H_{\mu\nu}^{(0)}(x)}{\partial k_\rho} \omega_\rho^{\ \rho'} k_{\rho'} +
\frac{1}{2} \frac{\partial^2 \hat H_{\mu\nu}^{(0)}(x)}{\partial k_\rho \partial k_\sigma} \omega_\rho^{\ \rho'} k_{\rho'} \omega_{\sigma}^{\ \sigma'} k_{\sigma'}+\cdots , \\
&\hat H_{\mu\nu}^{(1,L)\rho}(k_1,k_2,q) = \hat H_{\mu\nu}^{(1,L)\rho} (x_1,x_2) 
+ \frac{\partial \hat H_{\mu\nu}^{(1,L)\rho}(x_1,x_2)}{\partial k_{1\sigma}} \omega_\sigma^{\ \sigma'} k_{1\sigma'} 
+ \frac{\partial \hat H_{\mu\nu}^{(1,L)\rho}(x_1,x_2)}{\partial k_{2\sigma}} \omega_\sigma^{\ \sigma'} k_{2\sigma'} 
+\cdots ,
\end{align}
and so on, where $\omega_\rho^{\ \rho'}$ is a projection operator defined by $\omega_\rho^{\ \rho'} \equiv g_\rho^{\ \rho'} -\bar n_\rho n^{\rho'}$.\\

(2) Decompose the gluon field into longitudinal and transverse components, i.e.,
\begin{equation}
A_\rho(y) = A^+(y) \bar n_\rho + \omega_\rho^{\ \rho'} A_{\rho'}(y).
\end{equation}

(3) Apply the Ward identities such as, 
\begin{align}
& \frac{\partial \hat H^{(0)}_{\mu\nu}(x)}{\partial k_\rho} =  - \hat H^{(1)\rho}_{\mu\nu} (x,x),\\ 
& \frac{\partial \hat H^{(1,L)\rho}_{\mu\nu}(x_1,x_2)}{\partial k_{1,\sigma}}=  - \hat H^{(2,L)\rho\sigma}_{\mu\nu} (x_1,x_1,x_2)  - \hat H^{(2,M)\rho\sigma}_{\mu\nu}(x_1,x_1,x_2),\\
& p_\rho \hat H^{(1,L)\rho}_{\mu\nu}(x_1,x_2) = \frac{H^{(0)}_{\mu\nu} (x_1)}{x_2-x_1-i\epsilon}.\\
& p_\rho \hat H^{(2,L)\rho\sigma}_{\mu\nu}(x_1,x,x_2) = \frac{1}{x-x_1-i\epsilon} H^{(1,L)\sigma}_{\mu\nu} (x_1,x_2).\\
& p_\rho \hat H^{(2,M)\rho\sigma}_{\mu\nu}(x_1,x,x_2) = - \frac{1}{x_2-x_1-i\epsilon} H^{(1,L)\sigma}_{\mu\nu} (x_1,x_2) -\frac{1}{x_1-x + i\epsilon} H^{(1,R)\sigma}_{\mu\nu} (x,x_2).
\end{align}

(4) Add all terms with the same hard part together and we obtain, 
\begin{align}
&W_{\mu\nu}(q,p,S) = \sum_{j}\tilde{W}^{(j)}_{\mu\nu}(q,p,S),\\
&\tilde W_{\mu\nu}^{(0)}(q,p,S)=\frac{1}{2\pi} \int \frac{d^4k}{(2\pi)^4}
{\rm Tr}\big[\hat H_{\mu\nu}^{(0)}(x)\ \hat \Phi^{(0)}(k,p,S)\big],\label{eq:tW0}\\
 & \tilde W_{\mu\nu}^{(1)}(q,p,S) = \frac{1}{2\pi}  \int \frac{d^4k_1}{(2\pi)^4}\frac{d^4k_2}{(2\pi)^4} \sum_{c=L,R}
{\rm Tr}\bigl[\hat H_{\mu\nu}^{(1,c)\rho}(x_1,x_2) \omega_\rho^{\ \rho'} \hat \Phi^{(1)}_{\rho'}(k_1,k_2,p,S)\bigr], \label{eq:tW1}\\
& \tilde W_{\mu\nu}^{(2)}(q,p,S) = \frac{1}{2\pi}  \int \frac{d^4k_1}{(2\pi)^4} \frac{d^4k_2}{(2\pi)^4} \frac{d^4k}{(2\pi)^4} \sum_{c=L,R,M}
{\rm Tr}\bigl[\hat H_{\mu\nu}^{(2,c)\rho\sigma}(x_1,x_2,x) \omega_\rho^{\ \rho'} \omega_\sigma^{\ \sigma'}
\hat \Phi^{(2)}_{\rho'\sigma'}(k_1,k_2,k,p,S)\bigr],\label{eq:tW2}
\end{align}
where $\hat\Phi^{(j)}$'s are the gauge invariant un-integrated quark-quark and quark-$j$-gluon(s)-quark correlators given by,  
\begin{eqnarray}
&&\hat\Phi^{(0)}(k,p,S)=\int {d^4y} e^{iky} \langle p,S|\bar{\psi}(0){\cal{L}}(0;y)\psi(y)|p,S\rangle,  \label{eq:Phi0def}\\
&&\hat\Phi^{(1)}_\rho(k_1,k_2,p,S)
=\int {d^4y} {d^4z}
e^{ik_2z+ik_1(y-z)} \langle p,S|\bar\psi(0) {\cal L}(0;z)D_\rho(z){\cal L}(z;y)\psi(y)|p,S\rangle,  \label{eq:Phi1def}\\
&&\hat\Phi^{(2)}_{\rho\sigma}(k_1,k_2,k,p,S)=\int d^4yd^4y'd^4z e^{ik_1y + ik(y'-y) + ik_2(z-y')} \langle p,S|\bar\psi(0){\cal L}(0;y)
D_\rho(y) {\cal L}(y;y')D_\sigma(y'){\cal L}(y';z)\psi(z)|p,S\rangle,\label{eq:Phi2def}
\end{eqnarray}
$D(y)$ is the covariant derivative defined as $D_\rho(y) = -i\partial_\rho + g A_\rho(y)$.
The factor ${\cal{L}}(0;y)$ is obtained during summing different contributions with the same hard part together and  
is given by, 
\begin{align}
&{\cal{L}}(0;y)=\mathcal{L}^\dag(\infty; 0) \mathcal{L}(\infty; y),\label{CollinearGL} \\
& {\cal{L}}(\infty; y)= P e^{- i g \int_{y^-}^\infty d \xi^{-} A^+ ( y^+, \xi^-, \vec{y}_{\perp})}=1 - ig\int_{y^-}^{\infty} d\xi^-A^+ ( y^+, \xi^-, \vec{y}_{\perp}) + 
(-ig)^2\int_{y^-}^{\infty} d\xi^- \int_{\xi^-}^{\infty} d\eta^- A^+ ( y^+, \xi^-, \vec{y}_{\perp}) A^+ ( y^+, \eta^-, \vec{y}_{\perp}) + \cdots.
\end{align}
where $P$ stands for the path ordered integral. ${\cal{L}}(0;y)$ is nothing else but the well-known gauge link that makes 
the quark-quark or quark-$j$-gluon(s)-quark correlator, thus also the PDFs defined via them, gauge invariant.

In this way, we have constructed a theoretical framework for calculating the contributions to the hadronic tensor  
at the leading order (LO) in pQCD but leading as well as higher twists in a systematical way. 
The results are given in terms of the gauge invariant parton distribution and correlation functions (generally referred as PDFs).

We would like to emphasize in particular the following two further points derived directly from these expressions.
 
First, we note that after collinear expansion, the hard parts contained in the expressions 
for $\tilde W_{\mu\nu}^{(j)}$'s such as those given by Eqs.~(\ref{eq:tW0}-\ref{eq:tW2}) 
are only functions of the longitudinal component $x$. 
They are independent of other components of the parton momentum $k$.  
We can carry out the integration over these components of $k$'s and simplify them to,   
\begin{eqnarray}
&&\tilde W_{\mu\nu}^{(0)}(q,p,S)=\frac{1}{2\pi} \int {p^+dx}
{\rm Tr}\big[\hat H_{\mu\nu}^{(0)}(x)\ \hat \Phi^{(0)}(x,p,S)\big], \label{eq:tW0x}\\
&& \tilde W_{\mu\nu}^{(1)}(q,p,S)=\frac{1}{2\pi} \int {p^+ dx_1} {p^+dx_2} \sum_{c=L,R}
{\rm Tr}\bigl[\hat H_{\mu\nu}^{(1,c)\rho}(x_1,x_2) \omega_\rho^{\ \rho'}
\hat \Phi^{(1)}_{\rho'}(x_1,x_2,p,S)\bigr], \label{eq:tW1x}\\
&&\tilde W_{\mu\nu}^{(2)}(q,p,S)=\frac{1}{2\pi} \int p^+dx_1 p^+dx_2 p^+dx \sum_{c=L,R,M}
{\rm Tr}[\hat H_{\mu\nu}^{(2,c)\rho\sigma}(x_1,x_2,x)
\omega_\rho^{\ \rho'}\omega_\sigma^{\ \sigma'}
\hat\Phi^{(2)}_{\rho'\sigma'}(x_1,x_2,x,p,S)], \label{eq:tW2x}
\end{eqnarray}
where the matrix elements $\hat\Phi$'s are given by,
\begin{align}
&\hat\Phi^{(0)}(x,p,S) \equiv \int  \frac{d^4k}{(2\pi)^4}\delta(k^+-xp^+)\hat\Phi^{(0)}(k,p,S)  = \int \frac{dy^-}{2\pi} e^{ixp^+y^-} \langle p,S|\bar{\psi}(0){\cal{L}}(0;y^-)\psi(y^-)|p,S\rangle,  \label{eq:Phi0x}\\
&\hat\Phi^{(1)}_\rho(x_1,x_2,p,S) \equiv \int   \frac{d^4k_1}{(2\pi)^4} \frac{d^4k_2}{(2\pi)^4} \delta(k_1^+-x_1p^+)\delta(k_2^+-x_2p^+)\hat\Phi^{(1)}(k_1,k_2,p,S)\nonumber\\
&~~~~~~=\int \frac{dy^-}{2\pi} \frac{dz^-}{2\pi}  
e^{ix_2p^+z^-+ix_1p^+(y^--z^-)} \langle p,S|\bar\psi(0) {\cal L}(0;z^-)D_\rho(z^-){\cal L}(z^-;y^-)\psi(y^-)|p,S\rangle, \label{eq:Phi1x}\\
&\hat\Phi^{(2)}_{\rho\sigma}(x_1,x_2,x,p,S) \equiv \int   \frac{d^4k_1}{(2\pi)^4} \frac{d^4k_2}{(2\pi)^4} \frac{d^4k}{(2\pi)^4} \delta(k_1^+-x_1p^+)\delta(k_2^+-x_2p^+)\delta(k^+-xp^+) \hat\Phi^{(2)}(k_1,k_2,k,p,S)\nonumber\\
&~~~~~~=\int \frac{dy^-}{2\pi}\frac{d{y'}^-}{2\pi}\frac{dz^-}{2\pi} e^{ix_1p^+y^-+ixp^+({y'}^--y^-)+ix_2p^+(z^--{y'}^-)}  \langle p,S|\bar\psi(0){\cal L}(0;y^-)
D_\rho(y^-) {\cal L}(y^-;{y'}^-)D_\sigma({y'}^-){\cal L}({y'}^-;z^-)\psi(z^-)|p,S\rangle. 
\end{align}
From these expressions, we see explicitly that only $x_i$-dependences of the quark-quark and/or quark-$j$-gluon-quark correlators are involved. 
This means that only one dimensional imaging of the nucleon is relevant in inclusive DIS.
 
Second, due to the existence of the projection operator $\omega_\rho^{\ \rho'}$'s, the hard parts can be further simplified to a great deal. 
They are given by,
\begin{align}
&\hat H^{(0)}_{\mu\nu}(x)=\pi\hat h^{(0)}_{\mu\nu}\delta(x-x_B), \label{eq:H0simple}\\
&\hat H^{(1,L)\rho}_{\mu\nu}(x_1,x_2)\omega_\rho^{\ \rho'}
=\frac{\pi}{2q\cdot p}\hat h^{(1)\rho}_{\mu\nu}\omega_\rho^{\ \rho'}\delta(x_1-x_B), \label{eq:H1Lsimple}\\
&\hat H^{(2,L)\rho\sigma}_{\mu\nu}(x_1,x_2,x)\omega_\rho^{\ \rho'}\omega_\sigma^{\ \sigma'}=
\frac{2\pi}{(2q\cdot p)^2}\bigl[\bar n^\rho\hat h^{(1)\sigma}_{\mu\nu}+\frac{\hat N^{(2)\rho\sigma}_{\mu\nu}}{x_2-x_B-i\varepsilon}\bigr]
\omega_\rho^{\ \rho'}\omega_\sigma^{\ \sigma'}\delta(x_1-x_B), \label{eq:H2Lsimple}\\
& \hat H^{(2,M)\rho\sigma}_{\mu\nu}(x_1,x_2,x)\omega_\rho^{\ \rho'}\omega_\sigma^{\ \sigma'}=
\frac{2\pi }{(2q\cdot p)^2}\hat h^{(2)\rho\sigma}_{\mu\nu}
\omega_\rho^{\ \rho'}\omega_\sigma^{\ \sigma'}\delta(x-x_B), \label{eq:H2Msimple}
\end{align}
where $\hat h^{(0)}_{\mu\nu}=\gamma_\mu\slash{\hspace{-5pt}n}\gamma_\nu/p^+$,
$\hat h^{(1)\rho}_{\mu\nu}=\gamma_\mu\slash{\hspace{-5pt}\bar n}\gamma^\rho\slash{\hspace{-5pt}n}\gamma_\nu$,
$\hat h^{(2)\rho\sigma}_{\mu\nu}=p^+\gamma_\mu\slash{\hspace{-5pt}\bar n}\gamma^\rho
\slash{\hspace{-5pt} n}\gamma^\sigma\slash{\hspace{-5pt}\bar n}\gamma_\nu/2$  and
$\hat N^{(2)\rho\sigma}_{\mu\nu}=q^-\gamma_\mu\gamma^\rho\slash{\hspace{-5pt}n}\gamma^\sigma\gamma_\nu$ 
are matrices independent of $x_i$'s.
We insert them into Eqs.(\ref{eq:tW0x}-\ref{eq:tW2x}) and obtain the simplified expressions for the hadronic tensor as,
\begin{align}
& \tilde W^{(0)}_{\mu\nu}(q,p,S)=\frac{1}{2}{\rm Tr}\bigl[\hat h^{(0)}_{\mu\nu}\hat\Phi^{(0)}(x_B)\bigr],  \label{eq:tW0simple} \\
& \tilde W^{(1,L)}_{\mu\nu}(q,p,S) =
\frac{1}{4q\cdot p}{\rm Tr}\bigl[\hat h^{(1)\rho}_{\mu\nu}\omega_\rho^{\ \rho'}\hat\varphi^{(1)}_{\rho'}(x_B)\bigr], \label{eq:tW1Lsimple} \\
& \tilde W^{(2,L)}_{\mu\nu}(q,p,S)=
\frac{1}{(2q\cdot p)^2}\left\{{\rm Tr}\bigl[\hat h^{(1)\rho}_{\mu\nu} \omega_\rho^{\ \rho'}\hat\phi^{(2L)}_{\rho'}(x_B)\bigr]
+{\rm Tr}\bigl[\hat N^{(2)\rho\sigma}_{\mu\nu}\omega_\rho^{\ \rho'}\omega_\sigma^{\ \sigma'}\hat\varphi^{(2L)}_{\rho'\sigma'}(x_B)\bigr]\right\}, \label{eq:tW2Lsimple} \\
& \tilde W^{(2,M)}_{\mu\nu}(q,p,S)=
\frac{1}{(2q\cdot p)^2}{\rm Tr}\bigl[\hat h^{(2)\rho\sigma}_{\mu\nu}\omega_\rho^{\ \rho'}\omega_\sigma^{\ \sigma'} \hat\varphi^{(2M)}_{\rho'\sigma'}(x_B)\bigr], \label{eq:tW2Msimple}
\end{align}
where, for explicitness, we omit $p,S$ in the arguments of the correlators. These correlators are defined as,
\begin{align}
\hat\varphi^{(1)}_\rho(x_1)&\equiv\int dx_2\hat\Phi^{(1)}_\rho(x_1,x_2,p,S)
=\int \frac{p^+dy^-}{2\pi}e^{ixp^+y^-}\langle p,S|\bar\psi(0) D_\rho(0){\cal L}(0;y^-)\psi(y^-)|p,S\rangle, \label{eq:varphi1} \\
\hat\varphi^{(2M)}_{\rho\sigma}(x)&\equiv \int dx_1dx_2 \hat\Phi^{(2)}_{\rho\sigma}(x_1,x_2,x,p,S)
=\int \frac{p^+dy^-}{2\pi}e^{ixp^+y^-}
\langle p,S|\bar\psi(0) D_\rho(0){\cal L}(0;y^-)D_\sigma(y^-)\psi(y^-)|p,S\rangle, \label{eq:varphi2M} \\
\hat\varphi^{(2L)}_{\rho\sigma}(x_1)& \equiv \int \frac{dx dx_2}{x_2-x_1-i\varepsilon}\hat\Phi^{(2)}_{\rho\sigma}(x_1,x_2,x,p,S)\nonumber\\
&=\int \frac{dx_2}{x_2-x-i\epsilon} \frac{p^+dy^-}{2\pi}\frac{p^+dz^-}{2\pi}
e^{ixp^+y^- + i(x_2-x)p^+z^-} \langle p,S|\bar\psi(0){\cal L}(0;z^-) D_\rho(z^-)D_\sigma(z^-){\cal L}(z^-;y^-)\psi(y^-)|p,S\rangle, \label{eq:varphi2L} \\
\hat\phi^{(2L)}_\sigma(x_1)& \equiv \int dx dx_2 \bar n^\rho\hat\Phi^{(2)}_{\rho\sigma}(x_1,x_2,x,p,S)
=\int \frac{p^+dy^-}{2\pi}e^{ixp^+y^-} \langle p,S|\bar\psi(0) D^-(0)D_\sigma(0){\cal L}(0;y^-)\psi(y^-)|p,S\rangle. \label{eq:phi2L}
\end{align}
\end{widetext}
We see explicitly that all the involved components of the quark-$j$-gluon-quark correlators depends only on one single parton momentum.  
This means that only quark-$j$-gluon-quark correlators that depend on one single parton momentum are relevant in inclusive DIS.

We emphasize in particular that the results given by Eqs.~(\ref{eq:tW0}-\ref{eq:tW2})  and their simplified forms 
given by Eqs.~(\ref{eq:tW0simple}-\ref{eq:phi2L}) including the gauge links are derived in the collinear expansion. 
They are just the sum of the contributions from the diagram series shown in Fig.~1. 
This formalism provides us a basic theoretical framework for describing inclusive DIS at LO pQCD 
but leading and higher twist contributions in terms of gauge invariant PDFs.

The PDFs are defined in terms of QFT operators via these quark-quark correlators by expending them  
in terms of $\gamma$-matrices and basic Lorentz covariants. 
For example, for $\hat\Phi^{(0)}(x,p,S)$, we have,
\begin{align}
\hat\Phi^{(0)}(x)=\frac{1}{2}\Bigl[ & \Phi^{(0)}_S(x)+i\gamma_5\Phi^{(0)}_{PS}(x)+\gamma^\alpha\Phi^{(0)}_\alpha(x) \nonumber\\
&+\gamma_5\gamma^\alpha\tilde\Phi^{(0)}_\alpha(x)+i\sigma^{\alpha\beta}\gamma_5\Phi^{(0)}_{T\alpha\beta}(x)\Bigr].
\end{align}
The basic Lorentz covariants are constructed from $p_\alpha$, $n_\alpha$, $S_\alpha$ and $\varepsilon_{\alpha\beta\rho\sigma}$. 
We obtain the following general results,   
\begin{align}
 \Phi_S^{(0)}(x) =& M e(x), \label{eq:Phi0S}\\
 \Phi_{PS}^{(0)}(x) = &  \lambda M e_L(x), \label{eq:Phi0PS}\\
 \Phi_\alpha^{(0)}(x) = & p^+\bar{n}_\alpha f_1(x) + M \varepsilon_{\perp\alpha\rho}S_T^\rho f_T(x) + \frac{M^2}{p^+}n_\alpha f_3(x),\label{eq:Phi0V}\\
 \tilde\Phi_\alpha^{(0)}(x) = &  \lambda  p^+\bar{n}_\alpha g_{1L}(x) +  MS_{T\alpha} g_T(x)  +\lambda \frac{M^2}{p^+}n_\alpha g_{3L}(x),\label{eq:Phi0AV}\\
 \Phi_{T\rho\alpha}^{(0)}(x) = & p^+\bar{n}_{[\rho}S_{T\alpha]} h_{1T}(x)  - M\varepsilon_{\perp\rho\alpha} h(x) \nonumber\\
& + \lambda M\bar n_{[\rho}n_{\alpha]}  h_L(x) + \frac{M^2}{p^+}  n_{[\rho}S_{T\alpha]} h_{3T}(x),\label{eq:Phi0T}
\end{align}
where $\varepsilon_{\perp\rho\sigma} \equiv \varepsilon_{\alpha\beta\rho\sigma}\bar n^\alpha n^\beta$, 
and the anti-commutation symbol $A^{[\rho}B^{\sigma]} \equiv A^\rho B^\sigma - A^\sigma B^\rho$.
The scalar functions $f(x)$'s, $g(x)$'s and $h(x)$'s are the corresponding PDFs. 
There are totally 12 such functions, 3 of them, i.e. $f_1(x)$, $g_{1L}(x)$ and $h_{1T}(x)$, 
contribute at leading twist and have clear probability interpretations,
6 of them contribute at twist-3 and the other 3 contribute at twist-4. 
We further note that the three time reversal odd terms $e_L(x)$, $f_T(x)$ and $h(x)$ vanish in fact in the one dimensional case. 
We keep them in Eqs.~(\ref{eq:Phi0PS}-\ref{eq:Phi0T}) for late comparison with fragmentation functions. 

We also see that the PDFs involved here are all scale independent. 
This is because we have till now considered only the LO pQCD contributions, i.e. the tree diagrams. 
To go to higher order of pQCD, we take the loop diagrams, gluon radiations and so on into account. 
After proper handling of these contributions, we obtain the factorized form \cite{Collins:1989gx} 
where the PDFs acquire the scale $Q$-dependence governed by QCD evolution equations.
In practice, PDFs are parameterized and are given in the PDF library (PDFlib). 

In summary, for studying one dimensional imaging of the nucleon with inclusive DIS, we take the following steps.
\begin{itemize}
\item General symmetry analysis leads to the general form of the hadronic tensor and/or the cross section in terms of four independent structure functions.
\item Parton model without QCD interaction leads to LO in pQCD and leading twist results of structure functions in terms of 
$Q$-independent PDFs without (local) gauge invariance.
\item  Parton model with QCD multiple gluon scattering after collinear expansion leads to LO in pQCD, leading and 
higher twist contributions in terms of $Q$-independent but gauge invariant PDFs.  
\item Parton model with QCD multiple gluon scattering and ``loop diagram contributions" after collinear approximation, regularization and renormalization 
leads to leading and higher order pQCD, leading twist contributions 
in factorized forms in terms of $Q$-evolved and gauge invariant PDFs.
\end{itemize}

In the following, we will follow these four steps and summarize what we have achieved in the three dimensional case. 
As did in \cite{Liang:2015nia}, we will mainly focus on the theoretical framework at LO pQCD but taking leading and higher twist contributions 
into account consistently.
Before that, we would like to emphasize the following two of the historical developments that may be helpful to us in constructing 
the theoretical framework for the TMD case. 

First, as mentioned, the study of three dimensional imaging of the nucleon was triggered by 
the experimental observation of the single-spin left-right asymmetries (SSA) in the inclusive 
hadron-hadron collision with transversely polarized projectile or target. 
It was known that pQCD leads to negligibly small asymmetry for the hard part~\cite{Kane:1978nd}
but the observed asymmetry can be as large as 40\%~\cite{Adams:1991cs}. 
The hunting for such large asymmetries lasts for decades with the following milestones:
\begin{itemize}
\item In 1991, Sivers introduced~\cite{Sivers:1989cc} the asymmetric quark distribution in a transversely polarized nucleon 
that is now known as the Sivers function. 
\item In 1993, Boros, Liang and Meng proposed~\cite{Boros:1993ps} a phenomenological model 
that provides an intuitive physical picture showing that 
the asymmetry arises from the orbital angular momenta of quarks and what they called ``surface effect'' caused by the initial or final state interactions. 
\item In 1993, Collins published~\cite{Collins:1992kk} his proof that Sivers function has to vanish due to parity and time reversal invariance. 
\item In 2002, Brodsky, Hwang and Schmidt calculated~\cite{Brodsky:2002cx} SSA for SIDIS using an explicit example 
where they took the orbital angular momentum of quark and the multiple gluon scattering into account.
\item In 2002, immediately after \cite{Brodsky:2002cx}, Collins pointed out~\cite{Collins:2002kn} that 
the multiple gluon scattering is contained in the gauge link 
and that the conclusion of his proof in 1993 was incorrect because he forgot the gauge link. 
He further showed that by taking the gauge link into account the same proof leads to the conclusion that Sivers function for 
DIS and that for Drell-Yan have opposite sign. 
Belitsky, Ji and Yuan resolved~\cite{Ji:2002aa,Belitsky:2002sm} the problem of defining the gauge link 
for a TMD parton density in light-cone gauge where the gauge potential does not vanish asymptotically. 
\end{itemize}

The second historical development that we would like to mention concerns the azimuthal asymmetry study in SIDIS. 
It was shown by Georgi and Politzer in 1977~\cite{Georgi:1977tv} that final state gluon radiations 
lead to azimuthal asymmetries and could be used as a ``clean test to pQCD". 
However, soon after, in 1978, it was shown by Cahn~\cite{Cahn:1978se} that similar asymmetries can also 
be obtained if one includes intrinsic transverse momenta of partons. 
The latter, now named as Cahn effect, though power suppressed i.e. at higher twist, can be quite significant 
and can not be neglected since the values of the asymmetries themselves are usually not very large. 

The lessons that we learned from these historical developments are in particular the following two points, i.e., when studying TMDs, 
\begin{itemize}
\item it is important to take the gauge link into account;
\item higher twist effects can be important.\\
\end{itemize}
Both of them demand that, to describe SIDIS in terms of TMDs, we need the proper QFT formulation rather than the intuitive parton model.\\


\section{TMDs Defined via Quark-Quark Correlator}

The TMD PDFs of quarks are defined via the TMD quark-quark correlator $\Phi^{(0)}(x,k_\perp;p,S)$ given by Eq.~(\ref{eq:Phi0def}) (after integration over $k^-$). 
A systematical study has been given in \cite{Goeke:2005hb} and a very comprehensive treatment can also be found in \cite{Mulders}.
Here, we first expand it in terms of $\gamma$-matrices and obtain 
a scalar, a pseudo scalar, a vector, an axial-vector and an anti-symmetric and space reflection odd tensor part, i.e., 
\begin{widetext}
\begin{align}
\hat \Phi^{(0)}(x,k_\perp;p,S) =& \frac{1}{2} \bigl[ \Phi_S^{(0)}(x,k_\perp;p,S)
+ i\gamma_5 \Phi_{PS}^{(0)}(x,k_\perp;p,S) + \gamma^\alpha \Phi_\alpha^{(0)}(x,k_\perp;p,S) \nonumber\\
& ~+ \gamma_5\gamma^\alpha \tilde\Phi_\alpha^{(0)}(x,k_\perp;p,S) 
+ i\sigma^{\alpha\beta}\gamma_5 \Phi_{T\alpha\beta}^{(0)}(x,k_\perp;p,S) \bigr].\ \ \ \label{PhiExpansion}
\end{align}
The operator expressions of these coefficients are given by the traces of the quark-quark correlator with the corresponding Dirac matrices. 
For example, for the vector component, we have, 
\begin{eqnarray}
&&\hspace{-5mm}\Phi_\alpha^{(0)}(x,k_\perp;p,S) = \frac{1}{2} {\rm Tr} \bigl[ \gamma_\alpha \hat \Phi^{(0)}(x,k_\perp;p,S) \bigr]
= \int dz^- d^2z_\perp e^{i(xp^+z^- - \vec k_\perp \cdot \vec z_\perp)} \langle p,S|\bar{\psi}(0){\cal{L}}(0;z)\frac{\gamma_\alpha}{2}\psi(z)|p,S\rangle. 
\end{eqnarray}

We then analyze the Lorentz structure of each part by expressing it in terms of possible ``basic Lorentz covariants" and scalar functions. 
From $\hat \Phi^{(0)}(x,k_\perp;p,S)$, we obtain the results as \cite{Goeke:2005hb}, 
\begin{align}
 \Phi_S^{(0)}(x,k_\perp;p,S) &= M \Bigl[ e(x,k_\perp) - \frac{\varepsilon_{\perp\rho\sigma}k_\perp^\rho S_T^\sigma}{M}e_T^\perp(x,k_\perp) \Bigr], \label{eq:tmdPhiS}\\
\Phi_{PS}^{(0)}(x,k_\perp;p,S)& = M \Bigl[ \lambda e_L(x,k_\perp) - \frac{k_\perp \cdot S_T}{M} e_T(x,k_\perp) \Bigr],\label{eq:tmdPhiPS}\\
\Phi_\alpha^{(0)}(x,k_\perp;p,S) &=  p^+\bar{n}_\alpha \Bigl[ f_1(x,k_\perp) - \frac{\varepsilon_{\perp\rho\sigma}k_\perp^\rho S_T^\sigma}{M}f_{1T}^\perp(x,k_\perp) \Bigr] 
+ k_{\perp\alpha} \Bigl[ f^\perp(x,k_\perp) - \frac{\varepsilon_{\perp\rho\sigma}k_\perp^\rho S_T^\sigma}{M}f_{T}^{\perp1}(x,k_\perp) \Bigr] \nonumber\\
&  + \varepsilon_{\perp\alpha\rho}k_\perp^\rho \Bigl[ \lambda f_L^\perp(x,k_\perp) - \frac{k_\perp \cdot S_T}{M}f_T^{\perp2}(x,k_\perp) \Bigr] + \frac{M^2}{p^+}n_\alpha \Bigl[ f_3(x,k_\perp) - \frac{\varepsilon_{\perp\rho\sigma}k_\perp^\rho S_T^\sigma}{M}f_{3T}^\perp(x,k_\perp) \Bigr],\label{eq:tmdPhiV}\\
 \tilde\Phi_\alpha^{(0)}(x,k_\perp;p,S) & = p^+\bar{n}_\alpha \Bigl[ \lambda g_{1L}(x,k_\perp) - \frac{k_\perp \cdot S_T}{M} g_{1T}^\perp(x,k_\perp) \Bigr] + MS_{T\alpha} g'_T(x,k_\perp) - \varepsilon_{\perp\alpha\beta}k_\perp^\beta g^\perp(x,k_\perp) \nonumber \\
& + k_{\perp\alpha} \Bigl[ \lambda g_{L}^\perp(x,k_\perp) - \frac{k_\perp \cdot S_T}{M} g_{T}^{\perp}(x,k_\perp) \Bigr] + \frac{M^2}{p^+}n_\alpha \Bigl[ \lambda g_{3L}(x,k_\perp) - \frac{k_\perp \cdot S_T}{M} g_{3T}(x,k_\perp) \Bigr],\label{eq:tmdPhiAV}\\
\Phi_{T\rho\alpha}^{(0)}(x,k_\perp;p,S) &=  p^+\bar{n}_{[\rho}S_{T\alpha]} h_{1T}(x,k_\perp) - \frac{p^+\bar{n}_{[\rho}\varepsilon_{\perp\alpha]\beta}k_\perp^\beta}{M} h_1^\perp(x,k_\perp) + \frac{p^+\bar{n}_{[\rho}k_{\perp\alpha]}}{M} \Bigl[ \lambda h_{1L}^\perp(x,k_\perp) - \frac{k_\perp \cdot S_T}{M} h_{1T}^\perp(x,k_\perp) \Bigr] \nonumber\\
& + S_{T[\rho}k_{\perp\alpha]} h_T^\perp(x,k_\perp) - M\varepsilon_{\perp\rho\alpha} h(x,k_\perp) + M\bar n_{[\rho}n_{\alpha]} \Bigl[ \lambda h_L(x,k_\perp) - \frac{k_\perp \cdot S_T}{M} h_T(x,k_\perp) \Bigr] \nonumber\\
& + \frac{M^2}{p^+} \Bigl\{ n_{[\rho}S_{T\alpha]} h_{3T}(x,k_\perp) + \frac{n_{[\rho}k_{\perp\alpha]}}{M} \Bigl[ \lambda h_{3L}^\perp(x,k_\perp) - \frac{k_\perp \cdot S_T}{M} h_{3T}^\perp(x,k_\perp) \Bigr] - \frac{n_{[\rho}\varepsilon_{\perp\alpha]\beta}k_\perp^\beta}{M} h_3^\perp(x,k_\perp) \Bigr\}.\label{eq:tmdPhiT}
\end{align}
These scalar functions are known as TMD PDFs. There are totally 32 such TMD PDFs. 
Among them, 8 contribute at leading twist and they all have clear probability interpretations 
such as the number density $f_1(x,k_\perp)$, the helicity distribution $g_{1L}(x,k_\perp)$, the transversity $h_{1T}(x,k_\perp)$, 
the Sivers function $f_{1T}^\perp(x,k_\perp)$, the Boer-Mulders function $h_1^\perp(x,k_\perp)$ etc.; 
16 contribute at twist-3 and the other 8 contribute at twist-4. 
We emphasize that they are all scalar functions of $x$ and $k_\perp$, i.e., depending on $x$ and $k_\perp^2$. 

If we integrate over $d^2k_\perp$, terms that the basic Lorentz covariants are odd in $k_\perp$ vanish. 
Eqs.~(\ref{eq:tmdPhiS}-\ref{eq:tmdPhiT}) just reduce to the corresponding Eqs.~(\ref{eq:Phi0S}-\ref{eq:Phi0T}). 
At the leading twist, only 3 of 8 survive, i.e. the number density $f_1(x)$, the helicity distribution $g_{1L}(x)$ and the transversity $h_{1T}(x)$. 

We show the leading twist TMD PDFs in table~\ref{tab:TMDPDF1}. 
Those twist-3 TMD PDFs are shown in table~\ref{tab:TMDPDF2}. 
In these tables, we show also the results for the case that ${\cal L}=1$ , i.e. if we neglect the multiple gluon scattering and simply 
take a nucleon as an ideal gas system consisting of quarks and anti-quarks (see e.g. \cite{Mulders}). 
We also note that the conventions used here have the following systematics: 
$f$, $g$, and $h$ are for unpolarized, longitudinally and transversely polarized quarks; 
the subscript $L$ or $T$ stands for longitudinally  or transversely polarized nucleon, 
and those with subscript $1$ for leading twist, without number for twist-3 and with $3$ are for twist-4; 
the $\perp$ in the superscript denotes that the corresponding basic Lorentz covariant is $k_\perp$ dependent.

\begin{table}
\caption{The 8 leading twist TMD PDFs defined via the quark-quark correlator. A $\times$ means that the corresponding 
term disappears upon integrating the quark-quark correlator over $d^2k_\perp$.}\label{tab:TMDPDF1}
\begin{tabular}{c@{\hspace{0.6cm}}l@{\hspace{0.6cm}}c@{\hspace{0.6cm}}c@{\hspace{0.6cm}}c@{\hspace{0.6cm}}l} \hline
\begin{minipage}[c]{1.5cm}quark\\ polarization \end{minipage}  & \begin{minipage}[c]{1.5cm} nucleon \\ polarization \end{minipage}  
& TMD PDFs & if  ${\cal L}=1$ & integrated over $\vec k_\perp$ & name \rule[-0.39cm]{0mm}{0.95cm} \\[0.12cm] \hline 
\begin{minipage}[t]{1.5cm}~\\[-0.1cm]$U$ \end{minipage} & \hspace{0.5cm} $U$ & $f_1(x,k_\perp)$ &        & $f_1(x)$  & number density \rule[-0cm]{0mm}{0.48cm} \\ [-0.1cm]
       & \hspace{0.5cm} $T$ & $f_{1T}^\perp(x,k_\perp)$ &0 & $\times$ & Sivers function \\[0.12cm]  \hline
\begin{minipage}[t]{1.5cm} ~\\[-0.1cm] $L$ \end{minipage} & \hspace{0.5cm} $L$  & $g_{1L}(x,k_\perp)$ &  & $g_{1L}(x)$ & Helicity distribution \rule[-0cm]{0mm}{0.48cm}\\ [-0.1cm]
 & \hspace{0.5cm} $T$ & $g_{1T}^\perp(x,k_\perp)$ &  & $\times$ & Worm-gear/Trans-helicity distribution \\ [0.12cm]\hline
 \begin{minipage}[t]{1.5cm}~\\ [0.2cm] $T$ \end{minipage}& \hspace{0.5cm} $U$ & $h_{1}^\perp(x,k_\perp)$ & 0 & $\times$ & Boer-Mulders function \rule[-0cm]{0mm}{0.48cm}\\[-0.45cm]
 & \hspace{0.4cm} $T(\parallel)$ & $h_{1T}(x,k_\perp)$ &  & $h_{1T}(x)$ & transversity distribution \\
 & \hspace{0.4cm} $T(\perp)$ & $h_{1T}^\perp(x,k_\perp)$ &  &  & pretzelosity  \\
 & \hspace{0.5cm} $L$ & $h_{1L}^\perp(x,k_\perp)$ &   & $\times$ & Worm-gear/longi-transversity \\ [0.12cm]\hline
\end{tabular}
\end{table}

\begin{table}
\caption{The 16 twist-3 TMD PDFs defined via the quark-quark correlator. 
A $\times$ means that the corresponding term disappears upon integrating the quark-quark correlator over $d^2k_\perp$.}\label{tab:TMDPDF2}
\begin{tabular}{p{1.5cm}@{\hspace{0.3cm}}l@{\hspace{0.6cm}}l@{\hspace{0.2cm}}l@{\hspace{0.2cm}}l@{\hspace{0.6cm}}l@{\hspace{0.2cm}}l@{\hspace{0.3cm}}l@{\hspace{0.7cm}}l@{\hspace{0.2cm}}l@{\hspace{0.5cm}}l} \hline
\begin{minipage}[c]{1.5cm}quark\\ polarization \end{minipage}  & \begin{minipage}[c]{1.5cm} nucleon \\ polarization \end{minipage}  
 &  \multicolumn{3}{l}{TMD PDFs}& \multicolumn{3}{l}{if  ${\cal L}=1$} & \multicolumn{3}{l}{integrated over $\vec k_\perp$} \rule[-0.39cm]{0mm}{0.95cm} \\[0.12cm]  \hline 
\vspace{0.1cm} \hspace{0.5cm}$U$ & \hspace{0.5cm} $U$ &$e(x,k_\perp)$, & $f^\perp(x,k_\perp)$ & &0, & $f_1(x,k_\perp)/x$ &  & $e(x)$, &$\times$ &\\ [-0.2cm]
       & \hspace{0.5cm} $T$ & $e_{T}^\perp(x,k_\perp)$, &$f_{T}^{\perp1}(x,k_\perp)$, &$f_{T}^{\perp2}(x,k_\perp)$ &0, &0, &0& $\times$ &$\times$&$\times$ \\ [0.1cm] \hline
\vspace{0.1cm} \hspace{0.5cm} $L$ & \hspace{0.5cm} $L$  & $e_{L}(x,k_\perp)$, &$g_{L}^\perp(x,k_\perp)$ && 0, &$g_1(x,k_\perp)/x$& & $\times$, &$\times$ & \\ [-0.2cm]
 & \hspace{0.5cm} $T$ & $e_{T}(x,k_\perp)$, &$g'_{T}(x,k_\perp)$, &$g_{T}^\perp(x,k_\perp)$ & 0, &0,&$g_{1T}(x,k_\perp)/x$& $\times$&\multicolumn{2}{l}{~~~$g_{T}(x)$} \\ [0.1cm] \hline
\vspace{0.5cm} \hspace{0.5cm} $T$ & \hspace{0.5cm} $U$ & $h(x,k_\perp)$& & &{0}& & & $\times$& &\\[-0.6cm]
 & \hspace{0.4cm} $T(\parallel)$ & \multicolumn{3}{l}{$h_{T}^\perp(x,k_\perp)$} &\multicolumn{3}{l}{$h_{1T}^\perp(x,k_\perp)/x$} &\multicolumn{3}{l}{$\times$}\\[0.1cm]
 & \hspace{0.4cm} $T(\perp)$ &  \multicolumn{3}{l}{$h_{T}(x,k_\perp)$} &  \multicolumn{3}{l}{$h_{1T}(x,k_\perp)/x + k_\perp^2h_{1T}^\perp(x,k_\perp)/M^2x$}& \multicolumn{3}{l}{$\times$}  \\[0.1cm]
 & \hspace{0.5cm} $L$ &  \multicolumn{3}{l}{$h_{L}(x,k_\perp)$} & \multicolumn{3}{l}{$k_\perp^2h_{1L}^\perp(x,k_\perp)/M^2x$}& \multicolumn{3}{l}{$h_L(x)$}  \\[0.1cm] \hline
\hspace{0.5cm} $U$  & \hspace{0.5cm} $L$ &  \multicolumn{3}{l}{$f_{L}^\perp(x,k_\perp)$} &  \multicolumn{3}{l}{0}& \multicolumn{3}{l}{$\times$}  \\[0.1cm] 
\hspace{0.5cm} $L$  & \hspace{0.5cm} $U$ &  \multicolumn{3}{l}{$g^\perp(x,k_\perp)$} &  \multicolumn{3}{l}{0}& \multicolumn{3}{l}{$\times$}  \\[0.1cm] \hline
\end{tabular}
\end{table}

Higher twist TMD PDFs are also defined via quark-$j$-gluon(s)-quark correlators such as those given by Eqs.~(\ref{eq:varphi1}-\ref{eq:phi2L}).
Many of them are, however, not independent since they are related to those defined 
via the quark-quark correlator through the QCD equation of motion $\gamma \cdot D(z) \psi(z)=0$.
We can get the relations such as,
\begin{align}
& x\Phi_{\perp\rho}^{(0)}(x,k_\perp;p,S) = 
-\frac{n^\alpha}{p^+} \Bigl[ {\rm Re} \varphi_{\alpha\rho}^{(1)}(x,k_\perp;p,S)  
+ \varepsilon_{\perp\rho}^{~~~\sigma} {\rm Im} \tilde\varphi_{\alpha\sigma}^{(1)}(x,k_\perp;p,S) \Bigr],\\
& x\tilde\Phi_{\perp\rho}^{(0)}(x,k_\perp;p,S) = -\frac{n^\alpha}{p^+} \Bigl[ {\rm Re} \tilde\varphi_{\alpha\rho}^{(1)}(x,k_\perp;p,S)  
+ \varepsilon_{\perp\rho}^{~~~\sigma} {\rm Im} \varphi_{\alpha\sigma}^{(1)}(x,k_\perp;p,S) \Bigr].
\end{align}
It is interesting to see that~\cite{Song:2013sja}, although not generally proved, 
all the twist-3 TMD PDFs that are defined via quark-gluon-quark correlator $\varphi^{(1)}_\rho$ and involved in SIDIS 
are replaced by those defined via quark-quark correlator $\Phi^{(0)}$.

We would like to emphasize that fragmentation is just conjugate to parton distribution. 
A systematic study for the general structure of fragmentation function (FF) 
defined via the corresponding quark-quark correlator is presented in \cite{Chen:2015ora}.
We should have one to one correspondence between TMD PDFs and TMD FFs. 
E.g., corresponding to the quark-quark correlator $\Phi^{(0)}(k,p,S)$ given by Eq. (\ref{eq:Phi0def})
and the expanded form Eq. (\ref{PhiExpansion}), we have,
\begin{align}
\hat \Xi^{(0)} (k_F,p,S)=&\frac{1}{2\pi} \sum_X \int d^4 \xi e^{-ik_F\xi}
   \langle 0 | \mathcal{L}^\dagger(0,\infty)\psi(0) |hX\rangle \langle hX| \bar\psi(\xi) \mathcal{L}(\xi,\infty) |0\rangle. \label{eq:Xi0}
\end{align}
\begin{align}
\hat\Xi^{(0)}(z,k_{F\perp};p,S) =& \frac{1}{2} \bigl[ \Xi_S^{(0)}(z,k_\perp;p,S)
+ i\gamma_5 \Xi_{PS}^{(0)}(z,k_\perp;p,S) + \gamma^\alpha \Xi_\alpha^{(0)}(z,k_\perp;p,S)\nonumber\\
& ~+ \gamma_5\gamma^\alpha \tilde\Xi_\alpha^{(0)}(z,k_\perp;p,S) + i\sigma^{\alpha\beta}\gamma_5 \Xi_{T\alpha\beta}^{(0)}(z,k_\perp;p,S) \bigr]. \label{XiExpansion}
\end{align}
For spin-1/2 hadron, we have perfect one to one correspondence to those given by Eqs.~(\ref{eq:tmdPhiS}-\ref{eq:tmdPhiT}) for 
parton distributions in the nucleon, i.e.,  
\begin{align}
z\Xi_S^{(0)}(z,k_{F\perp}&;p,S) =M \Bigl[ E(z,k_{F\perp}) + \frac{\varepsilon_{\perp\rho\sigma}k_{F\perp}^\rho S_T^\sigma}{M}E_T^\perp(z,k_{F\perp}) \Bigr], \label{eq:tmdXiS}\\
 z\Xi_{PS}^{(0)}(z,k_{F\perp}&;p,S) =M \Bigl[ \lambda E_L(z,k_{F\perp}) + \frac{k_{F\perp} \cdot S_T}{M} E_T(z,k_{F\perp}) \Bigr],\label{eq:tmdXiPS}\\
 z\Xi_\alpha^{(0)}(z,k_{F\perp}&;p,S) = p^+\bar{n}_\alpha \Bigl[ D_1(z,k_{F\perp}) + \frac{\varepsilon_{\perp\rho\sigma}k_{F\perp}^\rho S_T^\sigma}{M}D_{1T}^\perp(z,k_{F\perp}) \Bigr] 
+ k_{F\perp\alpha} D^\perp(z,k_{F\perp}) + M\varepsilon_{\perp\alpha\rho}S_T^\rho D_{T}(z,k_{F\perp}), \nonumber\\
& + \varepsilon_{\perp\alpha\rho}k_{F\perp}^\rho \Bigl[ \lambda D_L^\perp(z,k_{F\perp}) + \frac{k_{F\perp} \cdot S_T}{M}D_T^{\perp}(z,k_{F\perp}) \Bigr] + \frac{M^2}{p^+}n_\alpha \Bigl[ D_3(z,k_{F\perp}) + \frac{\varepsilon_{\perp\rho\sigma}k_{F\perp}^\rho S_T^\sigma}{M}D_{3T}^\perp(z,k_{F\perp}) \Bigr],\label{eq:tmdXiV}\\
 z\tilde\Xi_\alpha^{(0)}(z,k_{F\perp}&;p,S) = p^+\bar{n}_\alpha \Bigl[ \lambda G_{1L}(z,k_{F\perp}) + \frac{k_{F\perp} \cdot S_T}{M} G_{1T}^\perp(z,k_{F\perp}) \Bigr] + MS_{T\alpha} G_T(z,k_{F\perp}) + \varepsilon_{\perp\alpha\beta}k_{F\perp}^\beta G^\perp(z,k_{F\perp}) \nonumber \\
& + k_{F\perp\alpha} \Bigl[ \lambda G_{L}^\perp(z,k_{F\perp}) + \frac{k_{F\perp} \cdot S_T}{M} G_{T}^{\perp}(z,k_{F\perp}) \Bigr] + \frac{M^2}{p^+}n_\alpha \Bigl[ \lambda G_{3L}(z,k_{F\perp}) + \frac{k_{F\perp} \cdot S_T}{M} G_{3T}(z,k_{F\perp}) \Bigr],\label{eq:tmdXiAV}\\
 z\Xi_{T\rho\alpha}^{(0)}(z,k_{F\perp} &;p,S) = p^+\bar{n}_{[\rho}S_{T\alpha]} H_{1T}(z,k_{F\perp}) + \frac{p^+\bar{n}_{[\rho}\varepsilon_{\perp\alpha]\beta}k_{F\perp}^\beta}{M} H_1^\perp(z,k_{F\perp}) + \frac{p^+\bar{n}_{[\rho}k_{F\perp\alpha]}}{M} \Bigl[ \lambda H_{1L}^\perp(z,k_{F\perp}) + \frac{k_{F\perp} \cdot S_T}{M} H_{1T}^\perp(z,k_{F\perp}) \Bigr] \nonumber\\
& + S_{T[\rho}k_{F\perp\alpha]} H_T^\perp(z,k_{F\perp}) + M\varepsilon_{\perp\rho\alpha} H(z,k_{F\perp}) + \bar n_{[\rho}n_{\alpha]} \Bigl[ M\lambda H_L(z,k_{F\perp}) + k_{F\perp} \cdot S_TH_T^{\prime\perp}(z,k_{F\perp}) \Bigr] \nonumber\\
& + \frac{M^2}{p^+} \Bigl\{ n_{[\rho}S_{T\alpha]} H_{3T}(z,k_{F\perp}) + \frac{n_{[\rho}\varepsilon_{\perp\alpha]\beta}k_{F\perp}^\beta}{M} H_3^\perp(z,k_{F\perp}) + \frac{n_{[\rho}k_{F\perp\alpha]}}{M} \Bigl[ \lambda H_{3L}^\perp(z,k_{F\perp}) + \frac{k_{F\perp} \cdot S_T}{M} H_{3T}^\perp(z,k_{F\perp}) \Bigr] \Bigr\}.\label{eq:tmdXiT}
\end{align}
Comparing them with the results given by Eqs. (\ref{eq:tmdPhiS}-\ref{eq:tmdPhiT}), 
we see clearly the one to one correspondence between FFs and PDFs. 
As an example, we show the 8 leading twist components in table~\ref{tab:TMDFF1}. 
We do not show the results for the case of ${\cal L}=1$ for FFs. 
This is because even if we neglect the multiple gluon scattering that leads to the gauge link, 
final state interactions can still exist between $h$ and $X$. 
In this case, time reversal invariance does not lead to zero results for the T-odd amplitudes.

\begin{table}
\caption{The 8 leading twist TMD FFs for spin-1/2 hadrons defined via the quark-quark correlator. 
A $\times$ means that the corresponding term disappears upon integrating the quark-quark correlator over $d^2k_{F\perp}$. }\label{tab:TMDFF1}
\begin{tabular}{p{1.8cm}l@{\hspace{0.6cm}}ccp{2.8cm}l} \hline
\begin{minipage}[c]{1.5cm}quark\\ polarization \end{minipage}  & \begin{minipage}[c]{1.5cm} hadron \\ polarization \end{minipage}  
 & TMD FFs & \hspace{0.5cm} & integrated over $\vec k_{F\perp}$ &  name  \rule[-0.39cm]{0mm}{0.95cm} \\[0.12cm]  \hline 
\vspace{0.1cm} \hspace{0.5cm} $U$ & \hspace{0.5cm} $U$ & $D_1(z,k_{F\perp})$ &        & \hspace{0.5cm} $D_1(z)$  & number density \\ [-0.2cm]
       & \hspace{0.5cm} $T$ & $D_{1T}^\perp(z,k_{F\perp})$ & & \hspace{0.6cm} $\times$ &  \\ [0.1cm] \hline
\vspace{0.1cm} \hspace{0.5cm} $L$ & \hspace{0.5cm} $L$  & $G_{1L}(z,k_{F\perp})$ &  & \hspace{0.5cm} $G_{1L}(z)$ & spin transfer (longitudinal) \\ [-0.2cm]
 & \hspace{0.5cm} $T$ & $G_{1T}^\perp(z,k_{F\perp})$ &  & \hspace{0.6cm} $\times$ &  \\ [0.1cm] \hline
\vspace{0.5cm} \hspace{0.5cm} $T$ & \hspace{0.5cm} $U$ & $H_{1}^\perp(z,k_{F\perp})$ &  & \hspace{0.6cm} $\times$ & Collins function \\[-0.6cm]
 & \hspace{0.4cm} $T(\parallel)$ & $H_{1T}(z,k_{F\perp})$ &  & \vspace{0.15cm} \hspace{0.5cm} $H_{1T}(z)$ & \vspace{-0.2cm} spin transfer (transverse)  \\[0.1cm]
 & \hspace{0.4cm} $T(\perp)$ & $H_{1T}^\perp(z,k_{F\perp})$ &  &  &   \\[0.1cm]
 & \hspace{0.5cm} $L$ & $H_{1L}^\perp(z,k_{F\perp})$ &  & \hspace{0.6cm} $\times$ &  \\[0.1cm] \hline
\end{tabular}
\end{table}

For spin-1 hadrons, the polarization is described by the polarization vector $S$ 
and also the polarization tensor $T$ (see e.g. \cite{Bacchetta:2000jk} and \cite{Chen:2015ora}). 
The tensor polarization part has five independent components. 
They are given by a Lorentz scalar $S_{LL}$, a Lorentz vector $S_{LT}^\mu=(0,S_{LT}^x,S_{LT}^y,0)$ 
and a Lorentz tensor $S_{TT}^{\mu\nu}$ that has two independent non-zero components 
$S_{TT}^{xx}$ and  $S_{TT}^{xy}$ in the rest frame of the hadron.  
These polarization parameters can be related to the probabilities for the particles in different spin states~\cite{Bacchetta:2000jk}. 
In this case, the TMD quark-quark correlator  $\hat \Xi^{(0)}(z,k_{F\perp};p,S)$ is decomposed into a spin independent part,
a vector polarization dependent part and a tensor polarization dependent part, i.e. $\hat \Xi^{(0)}(z,k_{F\perp};p,S)
=\hat \Xi^{U(0)}(z,k_{F\perp};p,S)+\hat \Xi^{V(0)}(z,k_{F\perp};p,S)+\hat \Xi^{T(0)}(z,k_{F\perp};p,S)$. 
The spin independent and vector polarization dependent part $\hat \Xi^{U+V(0)}(z,k_{F\perp};p,S)$ 
takes exactly the same decomposition as that for spin-1/2 hadron 
given by Eqs.~(\ref{eq:tmdXiS}-\ref{eq:tmdXiT}).  
The tensor polarization dependent part is presented in \cite{Chen:2015ora} and is given by, 
\begin{align}
z\Xi_S^{T(0)}(z,k_{F\perp}&;p,S) =M \Bigl[ S_{LL}E_{LL}(z,k_{F\perp}) + \frac{k_{F\perp} \cdot S_{LT}}{M} E_{LT}^\perp(z,k_{F\perp}) + \frac{k_{F\perp} \cdot S_{TT} \cdot k_{F\perp}}{M^2} E_{TT}^{\perp}(z,k_{F\perp})\Bigr], \label{eq:tmdXiTS}\\
 z\Xi_{PS}^{T(0)}(z,k_{F\perp}&;p,S) =M \Bigl[ \frac{\epsilon_\perp^{k_F S_{LT}}}{M} E_{LT}^{\prime\perp} (z,k_{F\perp}) + \frac{\epsilon_{\perp k_F \alpha} k_\beta S_{TT}^{\alpha\beta}}{M^2} E_{TT}^{\prime\perp}(z,k_{F\perp}) \Bigr],\label{eq:tmdXiTPS}\\
 z\Xi_\alpha^{T(0)}(z,k_{F\perp}&;p,S) = p^+\bar{n}_\alpha \Bigl[ S_{LL} D_{1LL}(z,k_{F\perp}) + \frac{k_{F\perp} \cdot S_{LT}}{M}D_{1LT}^\perp (z,k_{F\perp}) + \frac{k_{F\perp} \cdot S_{TT} \cdot k_{F\perp}}{M^2} D_{1TT}^\perp(z,k_{F\perp}) \Bigr] \nonumber\\
 & + k_{F\perp\alpha} \Bigl[ S_{LL}D_{LL}(z,k_{F\perp}) + \frac{k_{F\perp} \cdot S_{LT}}{M}D_{LT}^\perp (z,k_{F\perp}) + \frac{k_{F\perp} \cdot S_{TT} \cdot k_{F\perp}}{M^2} D_{TT}^\perp(z,k_{F\perp}) \Bigr] \nonumber\\
 & + MS_{LT\alpha} D_{LT}(z,k_{F\perp}) + k_{F\perp}^\rho S_{TT\rho\alpha} D_{TT}^{\prime\perp}(z,k_{F\perp}) \nonumber\\
 & + \frac{M^2}{p^+}n_\alpha \Bigl[ S_{LL} D_{3LL}(z,k_{F\perp}) + \frac{k_{F\perp} \cdot S_{LT}}{M}D_{3LT}^\perp (z,k_{F\perp}) + \frac{k_{F\perp} \cdot S_{TT} \cdot k_{F\perp}}{M^2} D_{3TT}^\perp(z,k_{F\perp}) \Bigr], \\
z\tilde\Xi_\alpha^{T(0)}(z,k_{F\perp}&;p,S) = p^+\bar{n}_\alpha \Bigl[ \frac{\varepsilon_\perp^{k_{F\perp}S_{LT}}}{M}G_{1LT}^\perp(z,k_{F\perp}) + \frac{\varepsilon_{\perp k_{F\perp}\rho}k_{F\perp\sigma}S_{TT}^{\rho\sigma}}{M^2}G_{1TT}^\perp(z,k_{F\perp}) \Bigr] \nonumber\\
 & + \varepsilon_{\perp\rho\alpha}k_{F\perp}^\rho \Bigl[ S_{LL}G_{LL}^\perp(z,k_{F\perp}) + \frac{k_{F\perp} \cdot S_{LT}}{M}G_{LT}^\perp (z,k_{F\perp}) + \frac{k_{F\perp} \cdot S_{TT} \cdot k_{F\perp}}{M^2} G_{TT}^\perp(z,k_{F\perp}) \Bigr] \nonumber\\
 & + M\varepsilon_{\perp\rho\alpha}S_{LT}^\rho G_{LT}(z,k_{F\perp}) + \varepsilon_{\perp\alpha\rho}k_{F\perp\sigma}S_{TT}^{\rho\sigma} G_{TT}^{\prime\perp}(z,k_{F\perp}) \nonumber\\
 & + \frac{M^2}{p^+}n_\alpha \Bigl[ \frac{\varepsilon_\perp^{k_{F\perp}S_{LT}}}{M}G_{3LT}^\perp(z,k_{F\perp}) + \frac{\varepsilon_{\perp k_{F\perp}\rho}k_{F\perp\sigma}S_{TT}^{\rho\sigma}}{M^2}G_{3TT}^\perp(z,k_{F\perp}) \Bigr] ,\\
 z\Xi_{T\rho\alpha}^{T(0)}(z,k_{F\perp} &;p,S) = \frac{p^+\bar{n}_{[\rho} \varepsilon_{\perp\alpha]\sigma}k_{F\perp}^\sigma}{M} \Bigl[ S_{LL} H_{1LL}^\perp(z,k_{F\perp}) + \frac{k_{F\perp} \cdot S_{LT}}{M}H_{1LT}^\perp (z,k_{F\perp}) + \frac{k_{F\perp} \cdot S_{TT} \cdot k_{F\perp}}{M^2} H_{1TT}^\perp(z,k_{F\perp}) \Bigr] \nonumber\\
 & + p^+\bar{n}_{[\rho} \varepsilon_{\perp\alpha]\sigma}S_{LT}^\sigma H_{1LT}(z,k_{F\perp}) + \frac{p^+\bar{n}_{[\rho} \varepsilon_{\perp\alpha]\sigma}k_{F\perp\delta}S_{TT}^{\sigma\delta}}{M} H_{1TT}^{\prime\perp}(z,k_{F\perp}) \nonumber\\
 & + M\varepsilon_{\perp\rho\alpha} \Bigl[ S_{LL} H_{LL}(z,k_{F\perp}) + \frac{k_{F\perp} \cdot S_{LT}}{M}H_{LT}^\perp (z,k_{F\perp}) + \frac{k_{F\perp} \cdot S_{TT} \cdot k_{F\perp}}{M^2} H_{TT}^\perp(z,k_{F\perp}) \Bigr] \nonumber\\
 & + \bar n_{[\rho}n_{\alpha]} \Bigl[ \varepsilon_\perp^{k_{F\perp}S_{LT}} H_{LT}^{\prime\perp}(z,k_{F\perp}) + \frac{\varepsilon_{\perp k_{F\perp}\sigma}k_{F\perp\delta}S_{TT}^{\sigma\delta}}{M}H_{TT}^{\prime\perp}(z,k_{F\perp}) \Bigr] \nonumber\\
 & + \frac{M^2}{p^+} \Bigl\{ \frac{n_{[\rho} \varepsilon_{\perp\alpha]\sigma}k_{F\perp}^\sigma}{M} \Bigl[ S_{LL} H_{3LL}^\perp(z,k_{F\perp}) + \frac{k_{F\perp} \cdot S_{LT}}{M}H_{3LT}^\perp (z,k_{F\perp}) + \frac{k_{F\perp} \cdot S_{TT} \cdot k_{F\perp}}{M^2} H_{3TT}^\perp(z,k_{F\perp}) \Bigr] \nonumber\\
 & \quad + n_{[\rho} \varepsilon_{\perp\alpha]\sigma}S_{LT}^\sigma H_{3LT}(z,k_{F\perp}) + \frac{n_{[\rho} \varepsilon_{\perp\alpha]\sigma}k_{F\perp\delta}S_{TT}^{\sigma\delta}}{M} H_{3TT}^{\prime\perp}(z,k_{F\perp}) \Bigr\}. \label{eq:tmdXiTT}
\end{align}

We see that, for the vector polarization dependent part, similar to nucleon TMD PDFs, 
there are totally 32 components, 8 contributes at leading twist, 16 at twist-3 and the other 8 at twist-4. 
For the tensor polarization dependent part, there are totally 40 components, 
where 10 contribute at leading twist, 20 at twist-3 and the other 10 at twist-4.
In Table~\ref{tab:TMDFFT1}, we list the twist-2 components for the tensor polarization dependent part.

\begin{table}
\caption{The 10 tensor polarization dependent TMD FFs for spin-1hadrons defined via the quark-quark correlator. 
A $\times$ means that the corresponding term disappears upon integrating the quark-quark correlator over $d^2k_{F\perp}$. }\label{tab:TMDFFT1}
\begin{tabular}{p{1.8cm}l@{\hspace{0.6cm}}lp{2.8cm}l} \hline
\begin{minipage}[c]{1.5cm}quark\\ polarization \end{minipage}  & \begin{minipage}[c]{1.5cm} hadron \\ polarization \end{minipage}  
 & TMD FFs  & integrated over $\vec k_{F\perp}$ &  name  \rule[-0.39cm]{0mm}{0.95cm} \\[0.12cm]  \hline 
\vspace{0.4cm} \hspace{0.5cm} $U$ & \hspace{0.5cm} $LL$ & $D_{1LL}(z,k_{F\perp})$    & \hspace{0.35cm} $D_{1LL}(z)$  & spin alignment \\ [-0.4cm]
      & \hspace{0.5cm} $LT$ & $D_{1LT}^\perp(z,k_{F\perp})$  & \hspace{0.6cm} $\times$ &  \\ [0.1cm]
      & \hspace{0.5cm} $TT$ & $D_{1TT}^\perp(z,k_{F\perp})$  & \hspace{0.6cm} $\times$ &  \\ [0.1cm] \hline
\vspace{0.1cm} \hspace{0.5cm} $L$ & \hspace{0.5cm} $LT$  & $G_{1LT}^\perp(z,k_{F\perp})$   & \hspace{0.6cm} $\times$ &  \\ [-0.2cm]
 & \hspace{0.5cm} $TT$ & $G_{1TT}^\perp(z,k_{F\perp})$  & \hspace{0.6cm} $\times$ &  \\ [0.1cm] \hline
\vspace{0.3cm} \hspace{0.5cm} $T$ & \hspace{0.5cm} $LL$ & $H_{1LL}^\perp(z,k_{F\perp})$  & \hspace{0.6cm} $\times$ &  \\[-0.4cm]
 & \hspace{0.5cm} $LT$ & $H_{1LT}(z,k_{F\perp}),~H_{1LT}^\perp(z,k_{F\perp})$  & \vspace{-0.2cm} \hspace{0.35cm} $H_{1LT}(z)$ & \vspace{-0.1cm}   \\[0.2cm]
 & \hspace{0.5cm} $TT$ & $H_{1TT}^\perp(z,k_{F\perp}),~H_{1TT}^{\prime\perp}(z,k_{F\perp})$  & \hspace{0.6cm}$\times$ &   \\[0.1cm]\hline
\end{tabular}
\end{table}

If we integrate over $d^2k_{F\perp}$, we have, corresponding to Eqs. (\ref{eq:tmdXiS}-\ref{eq:tmdXiT}), 
for the spin independent and vector polarization dependent part, 
\begin{align}
z\Xi_S^{U+V(0)}(z&;p,S) =M E(z), \label{eq:XiVS}\\
z\Xi_{PS}^{U+V(0)}(z&;p,S) = \lambda M E_L(z) ,\label{eq:XiVPS}\\
z\Xi_\alpha^{U+V(0)}(z&;p,S) = p^+\bar{n}_\alpha D_1(z) 
 + M \varepsilon_{\perp\alpha\rho} S_T^\rho D_{T}(z)
 + \frac{M^2}{p^+}n_\alpha D_3(z),\label{eq:XiVV}\\
z\tilde\Xi_\alpha^{U+V(0)}(z&;p,S) = \lambda p^+\bar{n}_\alpha  G_{1L}(z)  + MS_{T\alpha} G_T(z) + \lambda \frac{M^2}{p^+}n_\alpha  G_{3L}(z), \label{eq:XiVAV}\\
z\Xi_{T\rho\alpha}^{U+V(0)}(z&;p,S) = p^+\bar{n}_{[\rho}S_{T\alpha]} H_{1T}(z)  
  + M\varepsilon_{\perp\rho\alpha} H(z) + \lambda M \bar n_{[\rho}n_{\alpha]}  H_L(z)
 + \frac{M^2}{p^+}  n_{[\rho}S_{T\alpha]} H_{3T}(z). \label{eq:XiVT}
\end{align}
while for the tensor polarization dependent part, we have, 
\begin{align}
z\Xi_S^{T(0)}(z&;p,S) =M S_{LL}E_{LL}(z), \label{eq:XiTS}\\
z\Xi_{PS}^{T(0)}(z&;p,S) = 0,\label{eq:XiTPS}\\
z\Xi_\alpha^{T(0)}(z&;p,S) = p^+\bar{n}_\alpha S_{LL} D_{1LL}(z)
  + MS_{LT\alpha} D_{LT}(z,k_{F\perp}) +   \frac{M^2}{p^+}n_\alpha S_{LL} D_{3LL}(z), \\
z\tilde\Xi_\alpha^{T(0)}(z&;p,S) = M \varepsilon_{\perp\rho\alpha} S_{LT}^\rho G_{LT} (z),\\
z\Xi_{T\rho\alpha}^{T(0)}(z &;p,S) = p^+\bar{n}_{[\rho} \varepsilon_{\perp\alpha]\sigma}S_{LT}^\sigma H_{1LT}(z) 
  + M\varepsilon_{\perp\rho\alpha} S_{LL} H_{LL}(z)
  + \frac{M^2}{p^+}  n_{[\rho} \varepsilon_{\perp\alpha]\sigma} S_{LT}^\sigma H_{3LT}(z) . \label{eq:XiTT}
\end{align}
We see that, for the spin independent and vector polarization dependent parts, 12 components survive, 
3 of them contribute at twist-2, 6 at twist-3 and the other 3 at twist-4. 
This is exactly the same as those for PDFs for nucleon and we have exact one to one correspondence between the results given by Eqs.~(\ref{eq:XiVS}-\ref{eq:XiVT}) 
and those given by Eqs.~(\ref{eq:Phi0S}-\ref{eq:Phi0T}). 
 For the tensor polarization dependent part, there are only 8 components survive, 2 of them contribute at twist-2, 4 at twist-3 and the other 2 at twist-4. 
 This corresponds to the situation for PDFs for vector mesons. 
 We should have a one to one correspondence between the tensor polarization dependent FFs for 
 production of spin-1 hadron to those PDFs for spin-1 hadrons. We also listed the twist-2 components in table~\ref{tab:TMDFFT1}. 
 


\section{Accessing the TMDs in High Energy Reactions}

The TMDs can be studied in semi-inclusive high energy reactions such as 
SIDIS $e^-+N\to e^-+h+X$, 
semi-inclusive Drell-Yan $h+h\to l^++l^-+X$, and 
semi-inclusive hadron production in $e^+e^-$-annihilation $e^++e^-\to h_1+h_2+X$. 
With SIDIS, we study TMD PDFs and TMD FFs, 
while with Drell-Yan and $e^+e^-$ annihilation, we study TMD PDFs and TMD FFs separately. 
We now follow the same steps as those for inclusive DIS and briefly summarize what   
we already have in constructing the corresponding theoretical framework. 

\underline {(I) The general forms of hadronic tensors}: For all three classes of processes,  
the general forms of hadronic tensors have been studied and obtained. 
For SIDIS, it has been discussed in \cite{Gourdin:1973qx,Kotzinian:1994dv,Diehl:2005pc,Bacchetta:2006tn}
and it has been shown that one need 18 independent structure functions for spinless $h$. 
For Drell-Yan, a comprehensive study was made in \cite{Arnold:2008kf} 
and the number of independent structure functions is 48 for hadrons with spin 1/2. 
For $e^+e^-$-annihilation,  the study was presented in \cite{Pitonyak:2013dsu} and one needs 72 for spin-1/2 $h_1$ and $h_2$. 
The results are systematically presented in these papers and we will not repeat them here.
However, we would like to present as an example for the general form of the differential cross section for $e^-N\to e^-hX$. 
It is given by, 
\begin{align}
&\frac{d\sigma}{dxdydzd\psi d^2p_{h\perp}}  = \frac{\alpha_{em}^2}{xyQ^2} \Bigl(1+\frac{\gamma^2}{2x}\Bigr)
 \Bigl( {\cal F}_{UU} + \lambda_l {\cal F}_{LU} + \lambda {\cal F}_{UL} + \lambda_l \lambda {\cal F}_{LL} 
 + S_\perp {\cal F}_{UT} + \lambda_l S_\perp {\cal F}_{LT}\Bigr), \label{eq:SIDIS_CS} \\
&~~{\cal F}_{UU}= \frac{y^2}{1-\varepsilon}\Bigl( F_{UU,T}+\varepsilon F_{UU,L} +\sqrt{2\varepsilon(1+\varepsilon)} F_{UU}^{\cos\phi_h}\cos\phi_h +\varepsilon F_{UU}^{\cos2\phi_h} \cos2\phi_h\Bigr), \label{eq:SIDIS_UU}\\
&~~{\cal F}_{UL}= \frac{y^2}{1-\varepsilon}\Bigl( \sqrt{2\varepsilon(1+\varepsilon)} F_{UL}^{\sin\phi_h} \sin\phi_h+\varepsilon F_{UL}^{\sin2\phi_h} \sin2\phi_h\Bigr), \label{eq:SIDIS_UL}\\
&~~{\cal F}_{LU}=\frac{y^2}{1-\varepsilon} \sqrt{2\varepsilon(1-\varepsilon)}  F_{LU}^{\sin\phi_h} \sin\phi_h,\label{eq:SIDIS_LU}\\
&~~{\cal F}_{LL}= \frac{y^2}{1-\varepsilon}\Bigl( \sqrt{1-\varepsilon^2} F_{LL}+\sqrt{2\varepsilon(1-\varepsilon)}F_{LL}^{\cos\phi_h} \cos\phi_h\Bigr),\label{eq:SIDIS_LL}\\
&~~{\cal F}_{UT}=\frac{y^2}{1-\varepsilon}\Bigl[ \sqrt{2\varepsilon(1+\varepsilon)} F_{UT}^{\sin\phi_S} \sin\phi_S +  \bigl( F_{UT,T}^{\sin(\phi_h-\phi_S)} 
+\varepsilon F_{UT,L}^{\sin(\phi_h-\phi_S)}\bigr)\sin(\phi_h-\phi_S) \nonumber\\ 
&~~~~~~~~~~~~~~~~~~ +\varepsilon  F_{UT}^{\sin(\phi_h+\phi_S)} \sin(\phi_h+\phi_S) 
 +\sqrt{2\varepsilon(1+\varepsilon)}  F_{UT}^{\sin(2\phi_h-\phi_S)} \sin(2\phi_h-\phi_S)
 +\varepsilon F_{UT}^{\sin(3\phi_h-\phi_S)}\sin(3\phi_h-\phi_S)\Bigr], \label{eq:SIDIS_UT}\\
&~~{\cal F}_{LT}= \frac{y^2}{1-\varepsilon}\Bigl[ \sqrt{2\varepsilon(1-\varepsilon)} F_{LT}^{\cos\phi_S}\cos\phi_S  +\sqrt{1-\varepsilon^2}F_{LT}^{\cos(\phi_h-\phi_S)}\cos(\phi_h-\phi_S)+\sqrt{2\varepsilon(1-\varepsilon)}F_{LT}^{\cos(2\phi_h-\phi_S)}\cos(2\phi_h-\phi_S)\Bigr],\label{eq:SIDIS_LT}
\end{align}
where $\varepsilon=(1-y-\frac{1}{4}\gamma^2y^2)/(1-y+\frac{1}{2}y^2+\frac{1}{4}\gamma^2y^2)$, $\gamma=2Mx/Q$; 
the azimuthal angle $\psi$ is that of the out going lepton $\vec l'$ around the incident lepton beam with respect to an arbitrary fixed direction, 
which in case of transversely polarized target is taken as the direction of $\vec S_T$. 
In the deep inelastic limit, neglecting power suppressed terms, $d\psi=d\phi_S$.

From Eqs.~(\ref{eq:SIDIS_CS}-\ref{eq:SIDIS_LT}), we see explicitly that 
the 18 structure functions $F$'s are determined 
by the different azimuthal asymmetries in different polarized cases. 
These different azimuthal asymmetries are just defined by the average value of 
the corresponding trigonometric functions.
E.g., 
\begin{align}
&A_{UT}^{\sin(\phi_h-\phi_S)}=\langle \sin(\phi_h-\phi_S)\rangle_{UT}=
\frac{ F_{UT,T}^{\sin(\phi_h-\phi_S)}+\varepsilon F_{UT,L}^{\sin(\phi_h-\phi_S)}} {2(F_{UU,T}+\varepsilon F_{UU,L})},\\
&A_{UT}^{\sin(\phi_h+\phi_S)}=\langle \sin(\phi_h+\phi_S)\rangle_{UT}=
\frac{ \varepsilon F_{UT}^{\sin(\phi_h+\phi_S)}} {2(F_{UU,T}+\varepsilon F_{UU,L})}.
\end{align}

We also like to emphasize that they are the general forms independent of parton model and are valid at leading 
and higher twist and also leading and higher order in pQCD. 
 
\underline {(II) LO in pQCD and leading twist parton model results:}    
These are the simplest parton model results and can be obtained easily. 
E.g., for SIDIS, 
\begin{align}
&\frac{d\sigma^{(0)}}{dxdydzd\phi_S d^2p_{h\perp}}  = \frac{\alpha_{em}^2}{xyQ^2} \Bigl({\cal F}^{(0)}_{UU} + \lambda_l {\cal F}^{(0)}_{LU}  
+ \lambda {\cal F}^{(0)}_{UL} + \lambda_l \lambda {\cal F}^{(0)}_{LL} + S_\perp {\cal F}^{(0)}_{UT} + \lambda_l S_\perp {\cal F}^{(0)}_{LT}\Bigr), \label{eq:SIDIS_ltwist} \\
&~~{\cal F}^{(0)}_{UU}= A(y){\mathcal C}[f_1D_1]+2(1-y){\cal C}[w_1h_1^\perp H_1^\perp]\cos(2\phi_h),\label{eq:SIDIS_LT1}\\
&~~{\cal F}^{(0)}_{UL}= 2(1-y){\cal C}[w_1h_{1L}^\perp H_1^\perp] \sin(2\phi_h),\label{eq:SIDIS_LT2}\\
&~~{\cal F}^{(0)}_{LU}=0,\label{eq:SIDIS_LT4}\\
&~~{\cal F}^{(0)}_{LL}= C(y){\cal C}[g_{1L}D_1],\label{eq:SIDIS_LT5}\\
&~~{\cal F}^{(0)}_{UT}= A(y){\cal C}[w_2f_{1T}^\perp D_1] \sin(\phi_h-\phi_S) +2(1-y){\cal C}[w_3h_{1T} H_1^\perp] \sin(\phi_h+\phi_S)+2(1-y){\cal C}[w_4h_{1T}^\perp H_1^\perp] \sin(3\phi_h-\phi_S), \label{eq:SIDIS_LT6}\\
&~~{\cal F}^{(0)}_{LT}=  C(y){\cal C}[-w_2g_{1T}D_1]\cos(\phi_h-\phi_S), \label{eq:SIDIS_LT7}
\end{align}
where $A(y)=1+(1-y)^2$,  $C(y)=y(2-y)$, 
and ${\mathcal C}[w_i f D]$ denotes the convolution of $f$ and $D$ weighted by $w_i$, i.e., 
\begin{align}
{\cal C}[w_i f D]\equiv x\sum_q e_q^2 \int d^2k_\perp d^2k_{F\perp} \delta^{(2)}(k_\perp-k_{F\perp}-p_{hT}/z) w_i(k_\perp,k_{F\perp},p_{hT}) f^q(x,k_\perp)D^{q\to hX}(z,k_{F\perp}),
\end{align}
\end{widetext}
where the weights $w_i$'s are given by,
\begin{align}
w_1(\vec k_\perp,k_{F\perp})=&\frac{-2(\hat p_{hT}\cdot \vec k_{F\perp})(\hat p_{hT}\cdot \vec k_\perp) + (\vec k_\perp\cdot \vec k_{F\perp})}{MM_h},  \label{w1} \\
w_2(\vec k_\perp,k_{F\perp})=&-\frac{\hat p_{hT}\cdot \vec k_\perp}{M}, \label{w2}\\ 
w_3(\vec k_\perp,k_{F\perp})=&-\frac{\hat p_{hT}\cdot \vec k_{F\perp}}{M_h}, \label{w3}\\
w_4(\vec k_\perp,k_{F\perp})=&\frac{(\hat p_{hT}\cdot \vec k_\perp)(\vec k_{\perp}\cdot \vec k_{F\perp})+\vec k^2_\perp(\hat p_{hT}\cdot \vec k_{F\perp})}{M^2M_h} \nonumber\\
& -\frac{2(\hat p_{hT}\cdot \vec k_\perp)^2(\hat p_{hT}\cdot \vec k_{F\perp})}{M^2M_h}, \label{w4}  
\end{align}
where $\hat p_{hT}=\vec p_{hT}/|\vec p_{hT}|$ is the corresponding unit vector. 
The result can be obtained from those given e.g. in \cite{Bacchetta:2006tn} by neglecting all the power suppressed contributions. 

From Eqs.~(\ref{eq:SIDIS_ltwist}-\ref{eq:SIDIS_LT7}), we see in particular that, at leading twist, 
there exist 6 non-zero azimuthal asymmetries in different polarized cases, i.e., 
\begin{align}
&\langle\cos2\phi_h\rangle_{UU}^{(0)}=\frac{(1-y)}{A(y)}\frac{{\cal C}[w_1h_1^\perp H_1^\perp]}{{\mathcal C}[f_1D_1]}, \\
&\langle\sin2\phi_h\rangle_{UL}^{(0)}=\frac{(1-y)}{A(y)}\frac{{\cal C}[w_1h_{1L}^\perp H_1^\perp]}{{\mathcal C}[f_1D_1]},\\
&\langle\sin(\phi_h-\phi_S)\rangle_{UT}^{(0)}=\frac{{\cal C}[w_2f_{1T}^\perp D_1]}{2{\mathcal C}[f_1D_1]},\\
&\langle\sin(\phi_h+\phi_S)\rangle_{UT}^{(0)}=\frac{(1-y)}{A(y)}\frac{{\cal C}[w_3h_{1T} H_1^\perp]}{{\mathcal C}[f_1D_1]},\\
&\langle\sin(3\phi_h-\phi_S)\rangle_{UT}^{(0)}=\frac{(1-y)}{A(y)}\frac{{\cal C}[w_4h_{1T}^\perp H_1^\perp]}{{\mathcal C}[f_1D_1]},\\
&\langle\cos(\phi_h-\phi_S)\rangle_{LT}^{(0)}=\frac{C(y)}{2A(y)}\frac{{\cal C}[-w_2g_{1T} D_1]}{{\mathcal C}[f_1D_1]},
\end{align}
and they are determined by Boer-Mulders function $h_1^\perp$ convoluted with Collins function $H_1^\perp$,
the Worm-gear (longi-transversity) $h_{1L}^\perp$ convoluted with Collins function $H_1^\perp$,
the Sivers function $f_{1T}^\perp$ convoluted with $D_1$, 
the transversity  $h_{1T}$ convoluted with Collins function $H_1^\perp$, 
the Worm-gear (trans-helicity distribution) $g_{1T}^\perp$ convoluted with Collins function $H_1^\perp$. 
The azimuthal asymmetries $A_{UT}^{\sin(\phi_h\mp\phi_S)}$ are 
due to Sivers and Collins effects and are often referred as Sivers asymmetry and Collins asymmetry respectively. 

We would like to emphasize that the results given by Eqs.~(\ref{eq:SIDIS_ltwist}-\ref{w4}) is a complete parton model result at LO in pQCD and leading twist. 
It can be used to extract the TMDs at this order. 
Any attempt to go beyond LO in pQCD or to consider higher twists needs to go beyond this expression. 

\underline {(III) LO in pQCD, leading and higher twist results:} 
For the semi-inclusive processes where only one hadron is involved, either in the initial or the final state,  
it has been shown\cite{Liang:2006wp,Song:2010pf,Song:2013sja,Wei:2013csa,Wei:2014pma} 
that the collinear expansion can be applied. 
Such processes include: semi-inclusive DIS $e^-+N\to e^-+q(jet)+X$, and $e^+e^-$-annihilation $e^++e^-\to h+\bar q(jet)+X$. 
By applying the collinear expansion, we have constructed the theoretical frameworks for these processes 
with which leading as well as higher twist contributions can be calculated in a systematical way to LO in pQCD. 
The complete results up to twist-3 have been obtained in Refs.\cite{Song:2013sja,Wei:2013csa,Wei:2014pma}. 
For polarized $e^-+N\to e^-+q(jet)+X$, 
the simplified expressions for the hadronic tensor are very similar to those for the inclusive DIS given by Eqs.~(\ref{eq:tW0simple}-\ref{eq:tW2Msimple}),
\begin{widetext}
\begin{align}
&\tilde W^{(0,si)}_{\mu\nu}(q,p,S,k_\perp) =\frac{1}{2}{\rm Tr}\bigl[\hat h^{(0)}_{\mu\nu}\hat\Phi^{(0)}(x_B,k_\perp)\bigr],  \label{eq:tWsi0} \\
& \tilde W^{(1,L,si)}_{\mu\nu}(q,p,S,k_\perp) =
\frac{1}{4q\cdot p}{\rm Tr}\bigl[\hat h^{(1)\rho}_{\mu\nu}\omega_\rho^{\ \rho'}\hat\varphi^{(1,L)}_{\rho'}(x_B,k_\perp)\bigr],
\label{eq:tWsi1L} \\
& \tilde W^{(2,L,si)}_{\mu\nu}(q,p,S,k_\perp)=
\frac{1}{(2q\cdot p)^2}\left\{{\rm Tr}\bigl[\hat h^{(1)\rho}_{\mu\nu}
\omega_\rho^{\ \rho'}\hat\phi^{(2L)}_{\rho'}(x_B,k_\perp)\bigr]
+{\rm Tr}\bigl[\hat N^{(2)\rho\sigma}_{\mu\nu}\omega_\rho^{\ \rho'}\omega_\sigma^{\ \sigma'}\hat\varphi^{(2L)}_{\rho'\sigma'}(x_B,k_\perp)\bigr]\right\},
\label{eq:tWsi2L} \\
& \tilde W^{(2,M,si)}_{\mu\nu}(q,p,S,k_\perp)=
\frac{1}{(2q\cdot p)^2}{\rm Tr}\bigl[\hat h^{(2)\rho\sigma}_{\mu\nu}\omega_\rho^{\ \rho'}\omega_\sigma^{\ \sigma'}
\hat\varphi^{(2M)}_{\rho'\sigma'}(x_B,k_\perp)\bigr],
\label{eq:tWsi2M}
\end{align}
and the complete results up to twist-3 are given by,
\begin{align}
& \frac{d\sigma}{dxdyd^2k_\perp} = \frac{2\pi\alpha_{em}^2 e_q^2}{Q^2y} \bigl({\cal W}_{UU} + \lambda_l {\cal W}_{LU} + S_\perp {\cal W}_{UT} + \lambda {\cal W}_{UL} + \lambda_l \lambda {\cal W}_{LL} + \lambda_l S_\perp {\cal W}_{LT}\bigr), \\
& ~~~{\cal W}_{UU}(x,k_\perp,\phi) = A(y) f_q(x,k_\perp) - \frac{2x|\vec k_\perp|}{Q} B(y)f_q^\perp(x,k_\perp)\cos\phi,\\
& ~~~{\cal W}_{LU}(x,k_\perp,\phi) = - \frac{2x|\vec k_\perp|}{Q} D(y)g^\perp(x,k_\perp)\sin\phi,\\
& ~~~{\cal W}_{UT}(x,k_\perp,\phi,\phi_S) = \frac{|\vec k_\perp|}{M} A(y) f_{1T}^\perp(x,k_\perp)\sin(\phi-\phi_S) \nonumber\\
& \phantom{XXXXXXXXXXXX} + \frac{2xM}{Q} B(y) \bigl\{ \frac{k_\perp^2}{2M^2} f_T^\perp(x,k_\perp)\sin(2\phi-\phi_S) - f_T(x,k_\perp)\sin\phi_S \bigr\} ,\\
& ~~~{\cal W}_{UL}(x,k_\perp,\phi) = - \frac{2x|\vec k_\perp|}{Q} B(y) f_L^\perp(x,k_\perp) \sin\phi ,\\
& ~~~{\cal W}_{LL}(x,k_\perp,\phi) = C(y) g_{1L}(x,k_\perp) - \frac{2x|\vec k_\perp|}{Q} D(y) g_L^\perp(x,k_\perp)\cos\phi,\\
& ~~~{\cal W}_{LT}(x,k_\perp,\phi,\phi_S) = \frac{|\vec k_\perp|}{M} C(y)g_{1T}^\perp(x,k_\perp)\cos(\phi-\phi_S) \nonumber\\
& \phantom{XXXXXXXXXXXX} - \frac{2xM}{Q} D(y) \bigl[ g_T(x,k_\perp)\cos\phi_S - \frac{k_\perp^2}{2M^2} g_T^\perp(x,k_\perp)\cos(2\phi-\phi_S) \bigr].
\end{align}
where $B(y)=2(2-y)\sqrt{1-y}$, $D(y) = 2y\sqrt{1-y}$.
For unpolarized $e^-+N\to e^-+q(jet)+X$, the results up to twist-4 have also been obtained~\cite{Song:2010pf},
\begin{align}
 \frac{d\sigma_{UU}}{dxdyd^2k_\perp} = \frac{2\pi\alpha_{em}^2 e_q^2}{Q^2y} &\Bigl\{ A(y) f_1(x,k_\perp) - 2B(y) \frac{|\vec k_\perp|}{Q}xf^\perp(x,k_\perp)\cos\phi 
 - 4(1-y)\frac{|\vec k_\perp|^2}{Q^2}x[\varphi_3^{(1)\perp}(x,k_\perp) - \tilde\varphi_3^{(1)\perp}(x,k_\perp)]\cos2\phi \nonumber\\
& + 8(1-y)\frac{2x^2M^2}{Q^2}f_3(x,k_\perp)-2A(y) \frac{|\vec k_\perp|^2}{Q^2}x[\varphi_3^{(2,L)\perp}(x,k_\perp) - \tilde\varphi_3^{(2,L)\perp}(x,k_\perp) ] \Bigr\}. \label{eq:twist4}
\end{align}
These results are expressed in terms of the gauge invariant TMD PDFs or FFs and 
can be used as the basis for measuring these TMDs via the corresponding process at the LO in pQCD. 

We would like in particular to draw the attention to the results for $e^++e^-\to h+\bar q(jet)+X$  for $h$ with different spins~\cite{Wei:2014pma}.
Here, for hadronic tensor, we obtain again very much similar formulae also for this process, e.g., corresponding to Eqs. (\ref{eq:tWsi0}-\ref{eq:tWsi2L}), we have, 
\begin{align}
&\tilde W_{\mu\nu}^{(0, si)} (q,p,S,k'_{\perp}|e^+e^-)
= \frac{1}{2} {\rm Tr}\left[\hat h_{\mu\nu}^{(0)} \hat \Xi^{(0)} (z_B, k'_{\perp}) \right], \label{eq:tWee0} \\
&\tilde W^{(1,L,si)}_{\mu\nu)}(q,p,S,k'_{\perp}|e^+e^-)
= -\frac{1}{4p\cdot q} {\rm Tr} \left[ \hat h^{(1)\rho}_{\mu\nu} \omega_{\rho}^{\ \rho'} \hat \Xi^{(1)}_{\rho'} (z_B, k'_{\perp}) \right] , \label{eq:tWee1L} \\
&\tilde W^{(2,L,si)}_{\mu\nu}(q,p,S,k'_{\perp}|e^+e^-)
= \frac{1}{4(p\cdot q)^2} {\rm Tr} \Bigl[ \hat h^{(1)\rho}_{\mu\nu} \omega_\rho^{\ \rho'}
\hat \Xi_{\rho'}^{(2B)} (z_B, k'_\perp) + \hat N_{\mu\nu}^{(2)\rho\sigma} \omega_\rho^{\ \rho'} \omega_\sigma^{\ \sigma'} \hat \Xi_{\rho'\sigma'}^{(2C)} (z_B, k'_\perp) \Bigr], \label{eq:tWee2L}\\
&\tilde W^{(2,M,si)}_{\mu\nu}(q,p,S,k'_{\perp}|e^+e^-)
= \frac{1}{4(p\cdot q)^2} {\rm Tr} \left[ \hat h^{(2)\rho\sigma}_{\mu\nu} \omega_\rho^{\ \rho'} \omega_\sigma^{\ \sigma'} \hat \Xi_{\rho'\sigma'}^{(2A)} (z_B, k'_\perp) \right]. \label{eq:tWee2M}
\end{align}
\end{widetext}
A complete twist-3 results for differential cross sections, azimuthal asymmetries, 
and polarizations have been obtained for hadrons with spin-0, 1/2 and 1 in \cite{Wei:2014pma}. 
We see in particular for spin-1 hadrons, tensor polarization is involved, even at the leading twist level, we have, 
for $e^+e^-$ annihilation at the $Z^0$-pole, 
\begin{align}
& S_{LL}^{(0)} (y,z,p_T) = \frac{\sum_q T_0^q(y)D_{1LL} (z,p_T)}{2\sum_q T_0^q(y) D_1(z,p_T)},\\
& S_{LT}^{n(0)}(y, z, p_T) = - \frac{2|\vec p_T|}{3zM} \frac{\sum_q P_q (y) T_0^q (y) G_{1LT}^\perp (z,p_T)}{\sum_q T_0^q (y) D_1 (z, p_T)}, \\
& S_{LT}^{t(0)}(y, z, p_T) = - \frac{2|\vec p_T|}{3zM} \frac{\sum_q T_0^q (y) D_{1LT}^\perp (z,p_T)}{\sum_q T_0^q (y) D_1 (z, p_T)},\\
& S_{TT}^{nn(0)} (y, z, p_T) = - \frac{2|\vec p_T|^2}{3M^2} \frac{\sum_q T_0^q (y) D_{1TT}^\perp (z, p_T)}{\sum_q T_0^q (y) D_1 (y,p_T)},\\
& S_{TT}^{nt(0)}(y, z, p_T) = \frac{2|\vec p_T|^2}{3M^2} \frac{\sum_q P_q (y) T_0^q (y) G_{1TT}^\perp (z, p_T)}{\sum_q T_0^q (y) D_1 (y,p_T)},
\end{align}
where $n$ and $t$ denote the two transverse directions of the produced vector meson, one is normal to and the other is inside to the production plane. 
The coefficient $T_0^q(y) = c_1^qc_1^e[(1-y)^2+y^2] - c_3^qc_3^e[1-2y]$, 
$c_1^e = (c_V^e)^2+(c_A^e)^2$ and $c_3^e = 2 c_V^e c_A^e$; 
and $y$ in this reaction is defined as $y\equiv l_1^+/k^+$. 
$P_q(y)=T_1^q(y)/T_0^q(y)$ is the polarization of the quark produced at the $Z^0$-decay and 
$T_1^q(y) = -c_3^q c_1^e [(1-y)^2+y^2] + c_1^q c_3^e [1-2y]$.
This is a situation that is much less explored till now and is worthwhile for many further studies.

For the above-mentioned three kinds of semi-inclusive processes, there are always two hadrons involved. 
Collinear expansion has not been proved how to apply for such processes. 
It is unclear how one can calculate leading and higher twist contributions in a systematical way.
Nevertheless, twist-3 calculations that have been carried out for these processes~\cite{Mulders:1995dh,Boer:1997mf,Lu:2011th,Chen:2015uqa}, 
practically in the following steps: 

(i) draw Feynman diagrams with multiple gluon scattering to the order of one gluon exchange, 

(ii) insert the gauge link in the correlator wherever needed to make it gauge invariant, 

(iii) carry out calculations to the order $1/Q$. 

Although not proved, it is interesting to see that the results obtained this way reduce exactly to those obtained 
in the corresponding simplified cases where collinear expansion is applied if we take the corresponding fragmentation functions as $\delta$-functions.

\underline {(IV) TMD factorization and evolution:}
To describe the semi-inclusive high energy reactions mentioned above in terms of QCD and parton model, 
TMDs are needed and the factorization theorem has to involve transverse momentum dependence.
TMD factorization theorem has been established at the leading twist for semi-inclusive processes~\cite{Collins:1981uk,Collins:1981uw,Collins:1984kg,Collins:1985ue,Ji:2004hz,Idilbi:2004vb,Ji:2004xq,Ji:2004wu}. 
TMD evolution theory is also developing 
very fast~\cite{Henneman:2001ev,Zhou:2008mz,Kang:2011mr,Aybat:2011zv,Aybat:2011ge,Anselmino:2012aa,Sun:2013hua,Ma:2013aca,Echevarria:2014xaa,Aidala:2014hva,Kang:2014zza,Echevarria:2014rua,Collins:2014jpa,Kang:2015msa}.
There was a dedicated overviews by Daniel Boer~\cite{Boer:2015ala} in Spin2014. 
There is a dedicated annual workshop series since 2012.  
We refer the interested readers to these talks and overviews. 
  

\section{Available Data and Parameterizations}

Experiments have been carried out for all three kinds of semi-inclusive reactions. 
The results are summarized e.g. in a number of plenary talks in Spin2014 by Marcin Stolarski and Armine Rostomyan~\cite{Stolarski,Rostomyan}. 
Here, we will just briefly summarize the main data available and then try to sort out the TMD parameterizations that we already have. 

For SIDIS, there are measurements carried out by 
HERMES Collaboration~\cite{Airapetian:1999tv, Airapetian:2004tw, Airapetian:2009ae, Airapetian:2010ds, Airapetian:2012yg} at DESY,  
COMPASS Collaboration~\cite{Alexakhin:2005iw,Ageev:2006da,Alekseev:2008aa,Alekseev:2010rw, Adolph:2012sn, Adolph:2012sp, Adolph:2014pwc,Adolph:2014zba} at CERN, 
CLAS~\cite{Avakian:2010ae, Aghasyan:2011ha} and Hall A Collaboration~\cite{Qian:2011py, Huang:2011bc, Zhang:2013dow, Zhao:2014qvx} at Jefferson Laboratory. 
We list these SIDIS experiments in table~\ref{tab:DISExps} and briefly summarize the results in the following.

At DESY, the first measurement on single-spin asymmetries for SIDIS with longitudinally polarized target was carried out by HERMES~\cite{Airapetian:1999tv} 
for production of charged pions; then for the first time with transversely polarized target in \cite{Airapetian:2004tw}. 
They found non zero Sivers and Collins asymmetries $\langle\sin(\phi_h-\phi_S)\rangle_{UT}$ and $\langle\sin(\phi_h+\phi_S)\rangle_{UT}$. 
Measurements have then also carried out for $\pi^0$ and Kaons~\cite{Airapetian:2009ae, Airapetian:2010ds} 
and also for azimuthal asymmetries $\langle\cos\phi_h\rangle_{UU}$ and $\langle\cos(2\phi_h)\rangle_{UU}$ in the unpolarized case~\cite{Airapetian:2012yg}.

At CERN, COMPASS has carried out measurements on the Sivers and Collins asymmetries in reactions with Deuteron or proton targets for production of 
charged hadrons, pions and Kaons~\cite{Alexakhin:2005iw,Ageev:2006da,Alekseev:2008aa,Alekseev:2010rw, Adolph:2012sn, Adolph:2012sp, Adolph:2014pwc,Adolph:2014zba}, 
and also $\langle\cos\phi_h\rangle_{UU}$ and $\langle\cos(2\phi_h)\rangle_{UU}$ in the unpolarized case~\cite{Adolph:2014pwc}.

At JLab, CLAS has carried out the measurements~\cite{Avakian:2010ae, Aghasyan:2011ha} on $\langle\sin(2\phi_h)\rangle_{UL}$ for pions with different charges 
and $\langle\sin\phi_h\rangle_{LU}$ for $\pi^0$. 
Hall A Collaboration has made the measurements~\cite{Qian:2011py, Huang:2011bc, Zhang:2013dow, Zhao:2014qvx} on Collins and Sivers asymmetries 
for $\pi^\pm$ and $K^\pm$,  $\langle\cos(\phi_h-\phi_s)\rangle_{LT}$ for $\pi^\pm$ and  $\langle\sin(3\phi_h-\phi_s)\rangle_{UT}$. 
They are all summarized in table~\ref{tab:DISExps}.
 
\begin{table}
\caption{Available measurements on azimuthal asymmetries in SIDIS} \label{tab:DISExps}
\begin{tabular}{l@{\hspace{0.2cm}}l@{\hspace{0.2cm}}l@{\hspace{0.2cm}}l}\hline
collaboration  & ~~reaction & asymmetries  &  ref.'s \rule[-0.39cm]{0mm}{0.95cm} \\[0.12cm] \hline 
HERMES & $e^+N\to e^+\pi^\pm X$ & $A_{UL}^{\sin\phi_h}$, \ $A_{UL}^{\sin2\phi_h}$ &  \cite{Airapetian:1999tv} \\ 
 &$e^+N\to e^+\pi^\pm X$ & $A_{Siv}$,  $A_{Coll}$ &  \cite{Airapetian:2004tw} \\ 
 & $e^+N\to e^+\pi^{\pm,0}(K^{\pm}) X$ & $A_{Siv}$ &  \cite{Airapetian:2009ae} \\
&$e^+N\to e^+\pi^{\pm,0}(K^{\pm}) X$  & $A_{Coll}$ &  \cite{Airapetian:2010ds} \\ 
& $e^+N\to e^+ \pi^{\pm}(K^{\pm})X$ & $A_{UU}^{\cos\phi_h}$, $A_{UU}^{\cos2\phi_h}$  &  \cite{Airapetian:2012yg}
 \\ \hline
COMPASS & $\mu^-~ ^6LiD\to \mu^-h^\pm X$ & $A_{Siv}$, $A_{Coll}$ &  \cite{Alexakhin:2005iw,Ageev:2006da}   \\
& $\mu^-~ ^6LiD\to \mu^- \pi{^\pm}(K^{\pm,0}) X$ & $A_{Siv}$, $A_{Coll}$ &  \cite{Alekseev:2008aa}   \\
& $\mu^-NH_3\to \mu^-h^\pm X$ & $A_{Siv}$,  $A_{Coll}$ &   \cite{Alekseev:2010rw}  \\
& $\mu^-NH_3\to \mu^-h^\pm X$ &  $A_{Coll}$ &   \cite{Adolph:2012sn}  \\
& $\mu^-NH_3\to \mu^-h^\pm X$ & $A_{Siv}$ &   \cite{Adolph:2012sp}  \\
& $\mu^- NH_3\to \mu^-\pi^\pm(K^{\pm,0}) X$ & $A_{Siv}$, $A_{Coll}$ &   \cite{Adolph:2014zba}  \\
& $\mu^-~ ^6LiD\to \mu^-h^\pm X$ & $A_{UU}^{\cos\phi_h}$, $A_{UU}^{\cos2\phi_h}$  &  \cite{Adolph:2014pwc}  \\  \hline
CLAS & $e^-p \to e^-\pi^{\pm,0} X$ & $A_{UL}^{\sin2\phi_h}$ &  \cite{Avakian:2010ae}   \\
& $e^-~p \to e^-\pi^{0} X$ & $A_{LU}^{\sin\phi_h}$ &   \cite{Aghasyan:2011ha}  \\ \hline
JLab Hall A & $e^-~^3He \to e^-\pi^\pm X$ & $A_{Siv}$, $A_{Coll}$ &  \cite{Qian:2011py}   \\
&  $e^-~^3He \to e^-\pi^\pm X$ & $A_{LT}^{\cos(\phi_h-\phi_S)}$ &  \cite{Huang:2011bc} \\
& $e^-~^3He \to e^-\pi^\pm X$ & $A_{UT}^{\sin(3\phi_h-\phi_S)}$  &  \cite{Zhang:2013dow}   \\
& $e^-~^3He \to e^- K^\pm X$ & $A_{Siv}$, $A_{Coll}$ &   \cite{Zhao:2014qvx}    \\
\hline
\end{tabular}
\end{table}

Besides the data from SIDIS, 
we have now also measurements on the azimuthal asymmetries in $e^+e^-\to\pi+\pi+X$ by 
Belle~\cite{Abe:2005zx, Seidl:2008xc, Vossen:2011fk} and Babar collaboration~\cite{TheBABAR:2013yha}, and also preliminary results from BES~\cite{Guan:2015oaa}.
For Drell-Yan, there are data available on azimuthal asymmetries in e.g. reactions using pion beam~\cite{Badier:1981zpc,Falciano:1986zpc,Guanziroli:1988zpc,Conway:1989prd}, 
and $pp$ or $pD$ collisions~\cite{Zhu:2006gx,Zhu:2008sj}.

Although the data are still far from abundant enough to give a precise control of the TMDs involved, there are already different 
sets of TMD parameterizations extracted from them. We briefly sort them out in the following.

The first part concerns what people called ``the first phase parameterizations", i.e. TMD parameterizations without QCD evolutions. 
Here, we have in particular the following results available. 
We emphasize once more that all the results including the figures are taken from these references 
\cite{Anselmino:2005nn,Anselmino:2007fs,Schweitzer:2010tt,Signori:2014kda,Anselmino:2013lza,
Efremov:2004tp,Collins:2005ie,Arnold:2008ap,Anselmino:2008sga,Vogelsang:2005cs,Bacchetta:2011gx,
Anselmino:2013vqa,Barone:2008tn,Barone:2009hw,Zhang:2008nu,Zhang:2008ez,Lu:2009ip}. 
The interested readers are referred to these references for more details.  

(1) Transverse momentum dependence: 
This is usually taken as~\cite{Anselmino:2005nn,Anselmino:2007fs,Schweitzer:2010tt,Signori:2014kda,Anselmino:2013lza} 
a Gaussian in a factorized form independent of the longitudinal variable $z$ or $x$, e.g., 
\begin{align}
&f_1(x,k_\perp)=f_1(x) e^{-\vec k_\perp^2/\langle\vec k_\perp^2\rangle}/\pi\langle\vec k_\perp^2\rangle, \\
&D_1(z,k_{F\perp})=D_1(z)e^{-\vec k_{F\perp}^2/\langle\vec k_{F\perp}^2\rangle}/\pi\langle\vec k_{F\perp}^2\rangle.
\end{align}
The width has been fitted, the form and flavor dependence etc. have been tested. 
The typical values of the fitted widths are e.g.~\cite{Anselmino:2005nn}, $\langle\vec k_\perp^2\rangle=0.25$GeV$^2$, $\langle\vec k_{F\perp}^2\rangle=0.20$GeV$^2$. 
Roughly speaking, this is a quite satisfactory fit.  
However, it has also been pointed out, e.g. in \cite{Signori:2014kda} for the TMD FF, 
that the Gaussian form seems to depend on the flavor and even on $z$,  
which means that it is only a zeroth order approximation. 

(2) Sivers function: All the data available from HERMES~\cite{Airapetian:2004tw, Airapetian:2009ae, Airapetian:2010ds},  
COMPASS~\cite{Alexakhin:2005iw,Ageev:2006da,Alekseev:2008aa,Alekseev:2010rw, Adolph:2012sp, Adolph:2014zba}, 
and JLab Hall A~\cite{Qian:2011py, Huang:2011bc,Zhao:2014qvx} on Sivers asymmetries in SIDIS for pions and Kaons have been used for the parameterization. 
The Sivers function is usually parameterized~\cite{Efremov:2004tp,Collins:2005ie,Arnold:2008ap,Anselmino:2005nn,Anselmino:2008sga,Vogelsang:2005cs,Bacchetta:2011gx}  
in the form of the number density $f_q(x,k_\perp)$ multiplied by an $x$-dependent factor ${\cal N}_q(x)$ and a $k_\perp$-dependent factor $h(k_\perp)$, i.e.,
\begin{equation}
\Delta^N f_q(x,k_\perp)=2{\cal N}_q(x) h(k_\perp) f_q(x,k_\perp),
\end{equation}
where ${\cal N}_q(x)$ is taken as a binomial function of $x$, 
\begin{equation}
{\cal N}_q(x)={\cal N}_q x^{\alpha_q}(1-x)^{\beta_q}{(\alpha_q+\beta_q)^{\alpha_q+\beta_q}}/{\alpha_q^{\alpha_q} \beta_q^{\beta_q}},
\end{equation}
and $h(k_\perp)$ is taken as a Gaussian,
\begin{equation}
h(k_\perp)=\sqrt{2e}({|\vec k_\perp|}/{M_1})e^{-\vec k_\perp^2/M_1^2}.
\end{equation}
Here the Sivers function $\Delta^N f_q(x,k_\perp)$ is defined via, 
\begin{equation}
f_{q/N^\uparrow}(x,k_\perp)=f_{q/N}(x,k_\perp)+\frac{1}{2} \Delta^N f_q(x,k_\perp)\vec S\cdot(\hat p\times \hat k_\perp),
\end{equation}
which is related to the Sivers function $f_{1T}^\perp(x,k_\perp)$ defined in Eq.~(\ref{eq:tmdPhiV}) by,
\begin{equation}
\Delta^N f_q(x,k_\perp)=-\frac{2|\vec k_\perp|}{M} f_{1T}^{\perp q}(x,k_\perp).
\end{equation}
There exist already different sets such as 
the Bochum~\cite{Efremov:2004tp,Collins:2005ie,Arnold:2008ap}, 
the Torino~\cite{Anselmino:2005nn,Anselmino:2008sga,Bacchetta:2011gx} 
and the Vogelsang-Yuan~\cite{Vogelsang:2005cs} fits. 
One thing seems to be clear that the Sivers function is nonzero for proton and it has different signs for $u$- and $d$-quark, as shown in Fig.2.


\begin{figure}[!ht]
\includegraphics[width=0.45\textwidth]{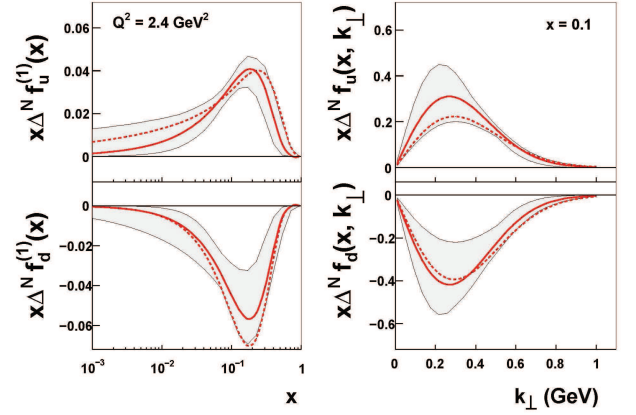}
\caption{Example of the parameterizations of the Sivers functions for $u$ and $d$ flavors at $Q^2 = 2.4 (GeV/c)^2$ by the Torino group. 
The figure is taken from \cite{Anselmino:2008sga}.}
\end{figure}

(3) Transversity and Collins function: 
A simultaneous extraction of them from SIDIS data from 
HERMES Collaboration~\cite{Airapetian:2004tw, Airapetian:2009ae, Airapetian:2010ds, Airapetian:2012yg} and   
COMPASS~\cite{Alexakhin:2005iw,Ageev:2006da,Alekseev:2008aa,Alekseev:2010rw, Adolph:2012sn, Adolph:2012sp, Adolph:2014zba}
on Collins asymmetries in SIDIS and $e^+e^-$ data of Belle~\cite{Abe:2005zx, Seidl:2008xc, Vossen:2011fk} 
have been carried out by the Torino group~\cite{Anselmino:2007fs,Anselmino:2013vqa}.
A similar form as that for the Sivers function has been taken, e.g., 
\begin{align}
&\Delta_T q(x,k_\perp)= \frac{1}{2}{\cal N}_q^T(x) [f_q(x)+\Delta q(x)] \nonumber\\
& \phantom{\Delta^N D_{h/q}(z,k_{F\perp})=} \times e^{-\vec k_\perp^2/\langle\vec k_\perp^2\rangle_T}/{\pi\langle\vec k_\perp^2\rangle},\\
& \Delta^N D_{h/q}(z,p_{hT})=2{\cal N}_q^C(z) D_{h/q}(z) h(p_{hT}) \nonumber\\
&  \phantom{\Delta^N D_{h/q}(z,p_{hT})=} \times  e^{-\vec p_{hT}^2/\langle\vec p_{hT}^2\rangle}/{\pi\langle\vec p_{hT}^2\rangle},\\
&{\cal N}_q^T(x)={\cal N}_q^T x^{\alpha}(1-x)^{\beta} \frac{(\alpha+\beta)^{\alpha+\beta}}{\alpha^{\alpha} \beta^{\beta}},\\
&{\cal N}_q^C(z)={\cal N}_q^C z^{\gamma}(1-z)^{\delta} \frac{(\gamma+\delta)^{\gamma+\delta}}{\gamma^{\gamma} \delta^{\delta}}.\\
&h(p_{hT})=\sqrt{2e}\frac{|\vec p_{hT}|}{M_h}e^{-\vec p_{hT}^2/M_h^2},
\end{align}
and it has been obtained that also the Collins function is nonzero and has different signs e.g. for $u\to \pi^+$ or $d\to \pi^+$, as shown in Fig.3.
Here, similar to the case for the Sivers function, the Collins function $\Delta^N D_{h/q}(z,k_{F\perp})$ is defined via,
\begin{equation}
D_{h/q^\uparrow}(z,p_{hT})=D_{q/N}(z,p_{hT})+\frac{1}{2} \Delta^N D_{h/q}(z,p_{hT})\vec s_q \cdot(\hat k_q\times \hat p_{hT}),
\end{equation}
which is related to the Collins function $H_{1T}^\perp(z,p_{hT})$ defined in Eq.~(\ref{eq:tmdXiV}) by,
\begin{equation}
\Delta^N D_{h/q}(z,p_{hT})=\frac{2|\vec p_{hT}|}{zM_h} H_{1T}^{\perp q}(z,p_{hT}).
\end{equation}


\begin{figure}[!ht]
\centerline{\psfig{figure=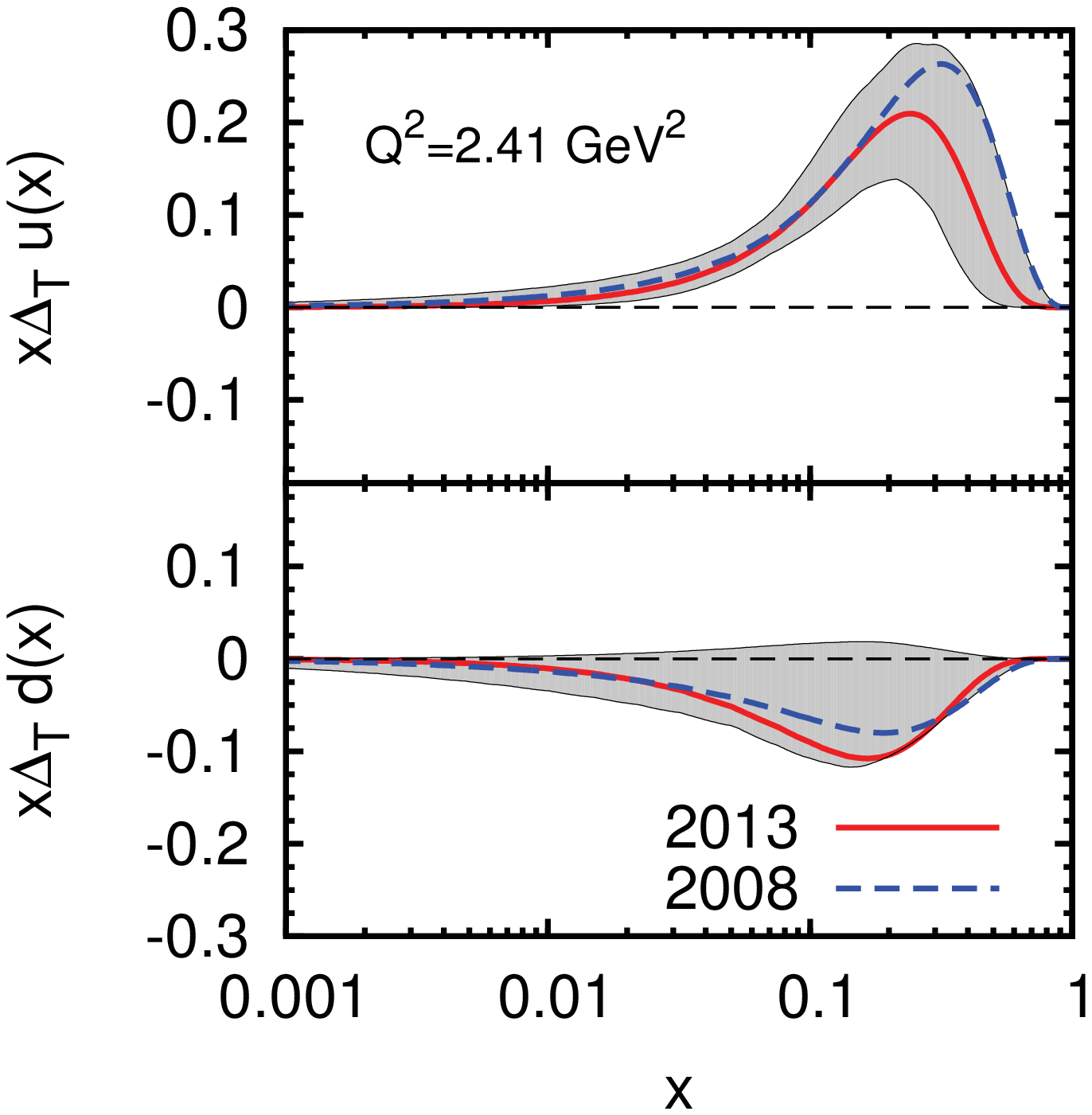, width=0.3\textwidth, angle=-90} \hspace{-2mm}\psfig{figure=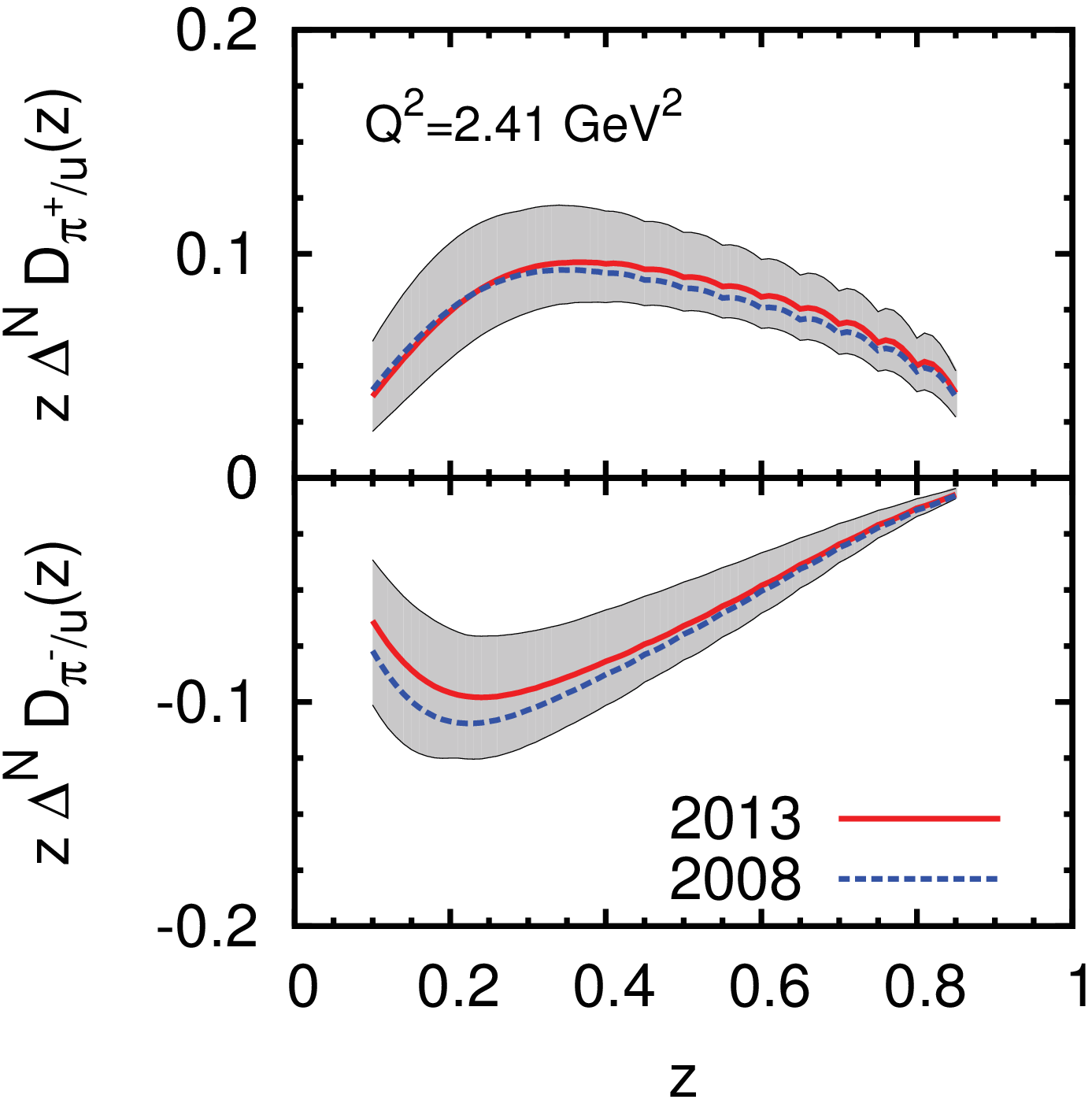, width=0.3\textwidth, angle=-90}}
\caption{Example of the Torino parameterizations of the transversity and Collins function. 
In the left panel, we see the transversities $x\Delta_Tq(x)=xh_{1q}(x)$ for $q = u,d$; in the right panel, we see the first moments of the favored and disfavored Collins functions. The figure is taken from \cite{Anselmino:2013vqa}.}
\end{figure}

(4) Boer-Mulders function: 
It was pointed out that~\cite{Barone:2009hw} the HERMES and COMPASS data 
on $\langle\cos2\phi\rangle$ asymmetry~\cite{Airapetian:2012yg,Adolph:2014pwc}
provide the first experimental evidence of the Boer-Mulders effect in SIDIS. 
Studies in this direction has been made in \cite{Barone:2008tn,Barone:2009hw} 
to extract Boer-Mulders function from the SIDIS data~\cite{Airapetian:2012yg,Adolph:2014pwc} 
and in \cite{Zhang:2008nu,Zhang:2008ez,Lu:2009ip} 
to extract from Drell-Yan data~\cite{Badier:1981zpc,Falciano:1986zpc,Guanziroli:1988zpc,Conway:1989prd,Zhu:2006gx,Zhu:2008sj}.
A fit to the first moments of Boer-Mulders function of $u$ and $d$ quark is shown in Fig. 4.
The form was taken again similar to the Sivers function, just multiply the Sivers function by a constant, e.g.,
\begin{equation}
 h_1^{\perp q}(x,k_\perp) = \lambda_q f_{1T}^{\perp q}(x,k_\perp).
 \end{equation} 
However, we would like to point out that the $\langle\cos2\phi\rangle$ asymmetry receives twist-4 contributions 
due to the Cahn effect~\cite{Cahn:1978se}. 
A proper treatment of such twist-4 effect involves twist-4 TMDs as shown in Eq.~(\ref{eq:twist4}) and in \cite{Song:2010pf}. 
Because of the multiple gluon scattering shown in Fig. 1, the twist-4 effects could be very much different 
from that given in \cite{Cahn:1978se} the results in which corresponds to the case of ${\cal L}=1$. 
A careful check might change the conclusion obtained in \cite{Zhang:2008nu,Zhang:2008ez,Barone:2008tn,Barone:2009hw,Lu:2009ip}.

 
\begin{figure}[!ht]
\centerline{ \hspace{5mm}\psfig{figure=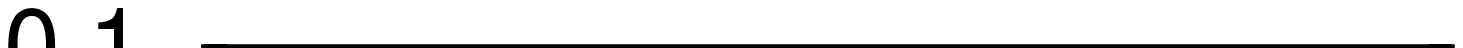, width=0.2\textwidth, angle=-90}} 
\caption{First extractions of the Boer-Mulders function $h_1^{\perp u}(x)$ and $h_1^{\perp d}(x)$. 
This figure is taken from Ref.\cite{Barone:2009hw}.}
\end{figure}

 Attempts to parameterize other TMDs such as pretzelocity $h^{\perp}_{1T}$ have also been made~\cite{Zhu:2011ir}. 
Although there is no enough data to give high accuracy constraints, the qualitative features obtained are also interesting.

\begin{figure}[!ht]
\centerline{\psfig{figure=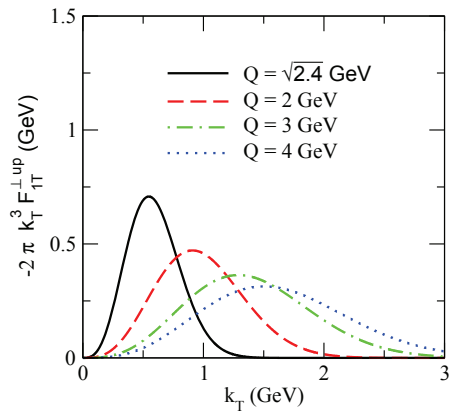, width=0.33\textwidth}} 
\caption{Example showing the TMD evolution of the Gaussian parameterization in the low $k_\perp$-region. 
The curves show the evolved Bochum Gaussian fits of up quark Sivers function at $x = 0.1$. 
This figure is taken from Ref.\cite{Aybat:2011ge}.}\label{fig:EvoGaussian}
\end{figure}

\begin{figure}[!ht]
\includegraphics[width=0.38\textwidth]{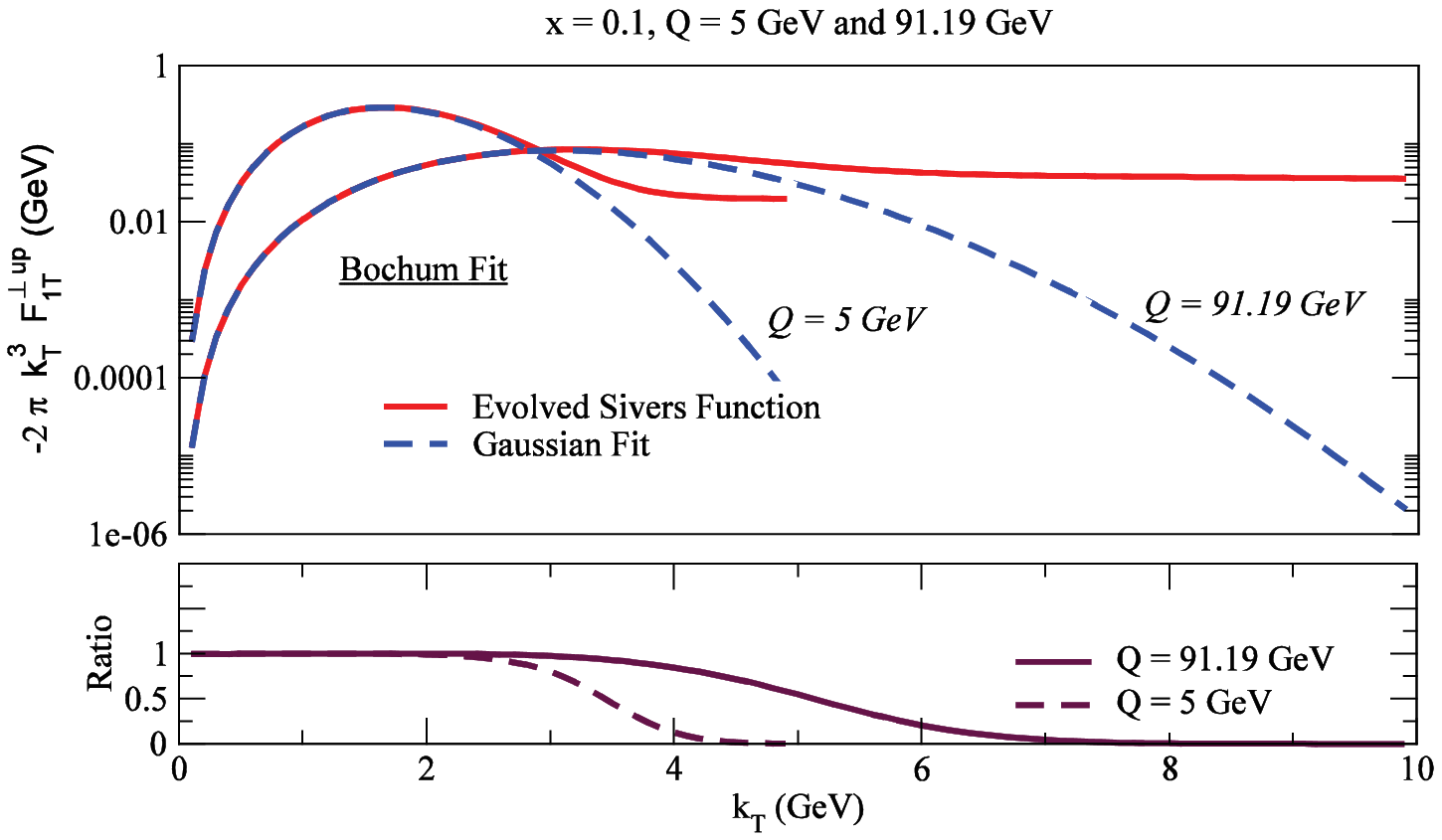}
\vspace{4mm}
\includegraphics[width=0.38\textwidth]{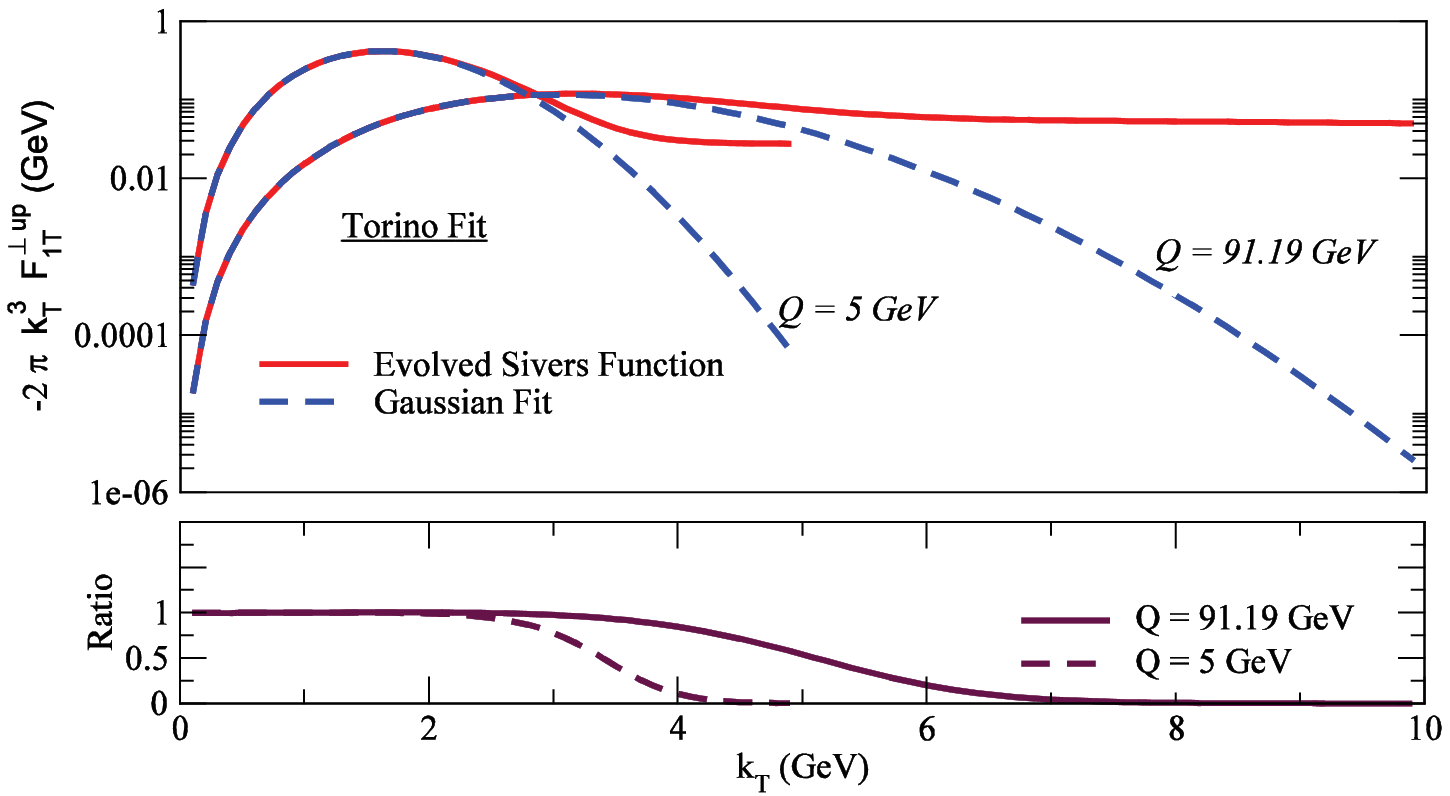}
\caption{Example showing the evolved $k_\perp$ dependence in the large $k_\perp$ region. 
Here we see the up-quark Sivers function at $Q=5$ GeV and $Q=91.19$ compared with the corresponding Gaussian fits at low-$k_\perp$ region at $x = 0.1$. 
This figure is taken from Ref.\cite{Aybat:2011ge}.}
\label{fig:EvoForm}
\end{figure}

\begin{figure}[!ht]
\centerline{\psfig{figure=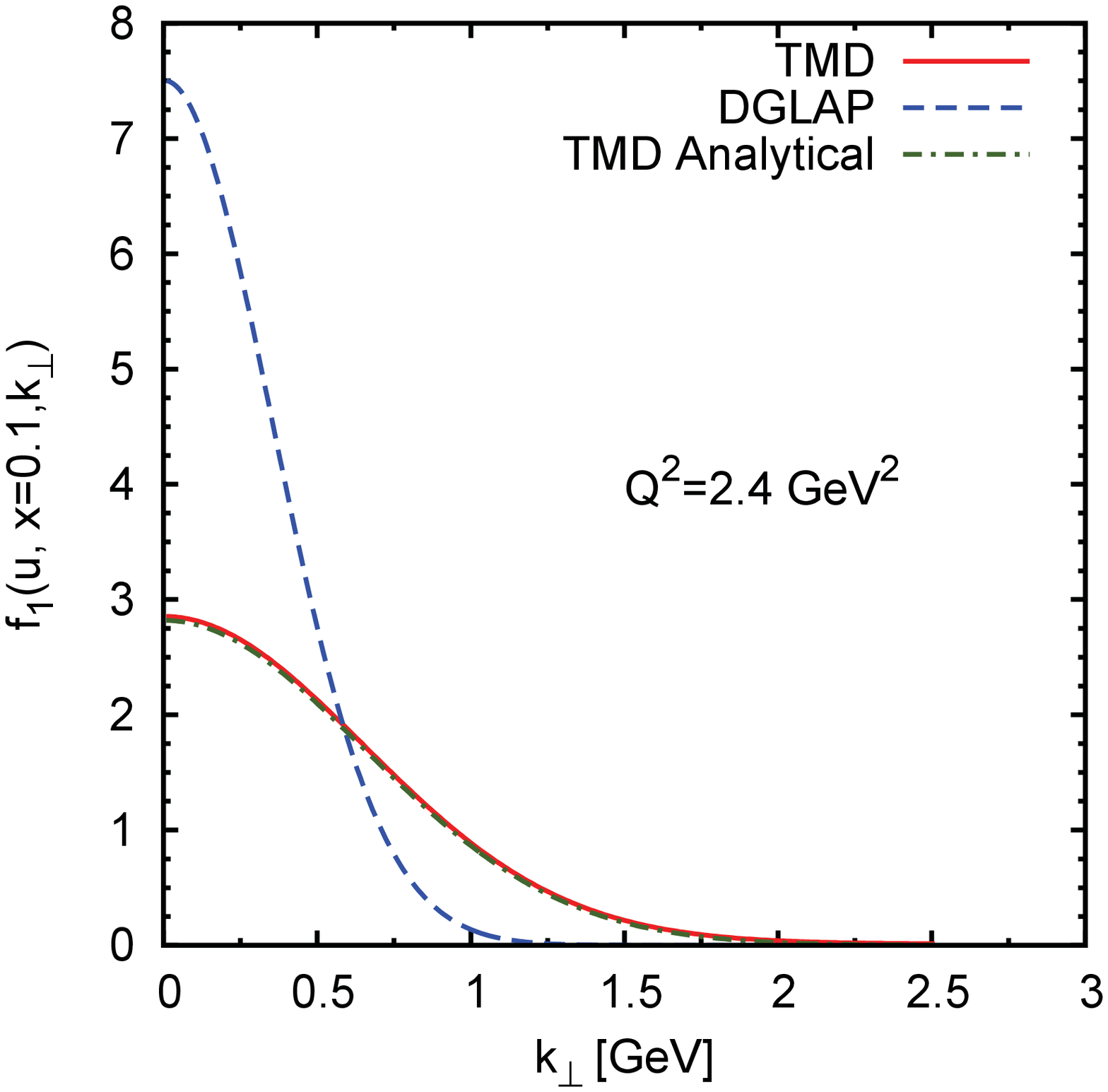, width=0.23\textwidth, angle=-90} \hspace{-22mm}\psfig{figure=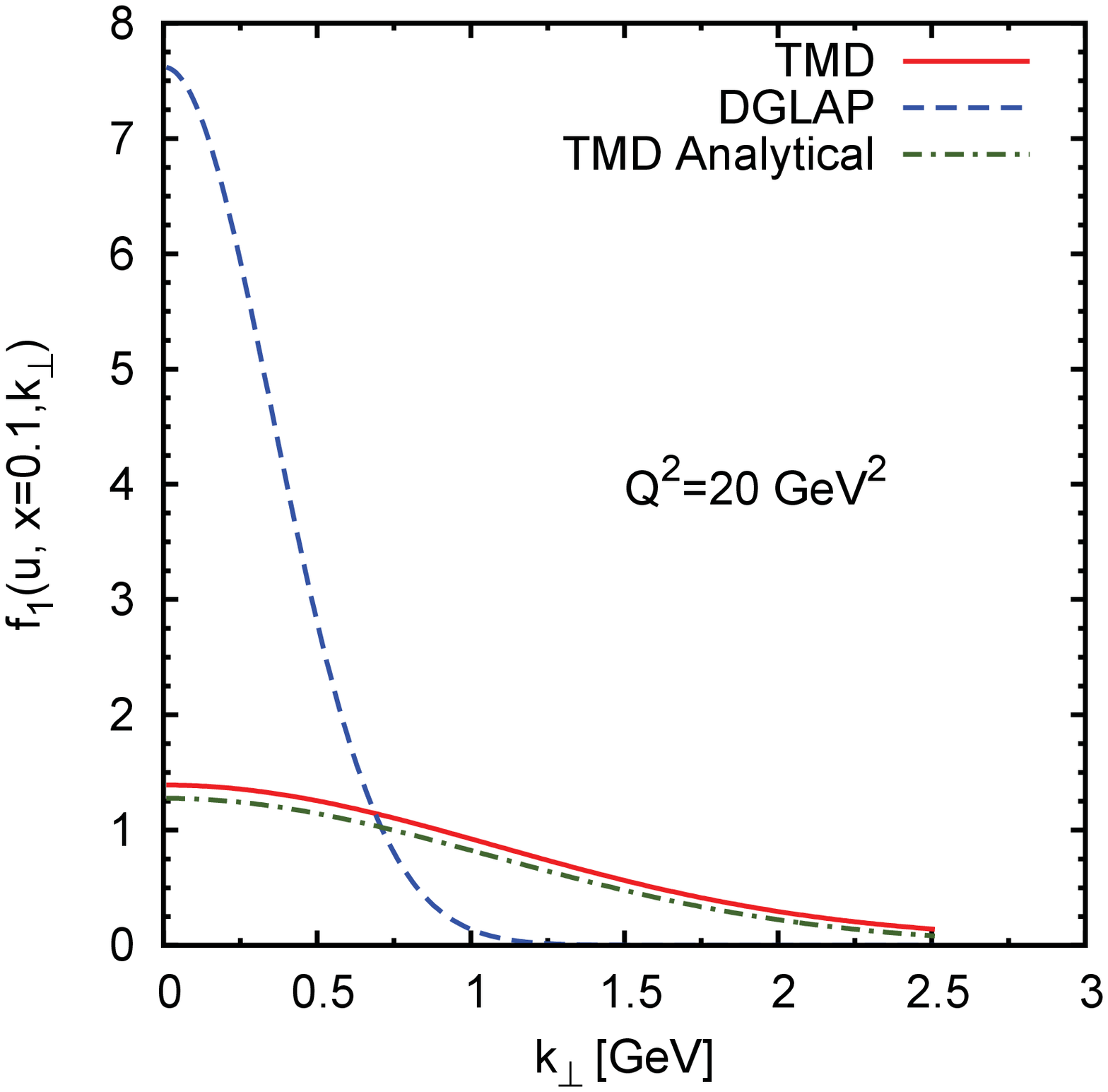, width=0.23\textwidth, angle=-90}}
\caption{Example showing the difference between the results of the TMD evolution with a DGLAP evolution for $x$-dependence only for unpolarized TMD PDF. 
This figure is taken from Ref.\cite{Anselmino:2012aa}.}
\label{fig:EvoDGLAP}
\end{figure}

The second part concerns the QCD evolution of the TMDs. 
As mentioned earlier, this is a topic that develops very fast recently. 
A partial list of recent dedicated publications is \cite{Henneman:2001ev,Zhou:2008mz,Kang:2011mr,Aybat:2011zv,Aybat:2011ge,Anselmino:2012aa,Sun:2013hua,Ma:2013aca,Echevarria:2014xaa,Aidala:2014hva,Kang:2014zza,Echevarria:2014rua,Collins:2014jpa,Kang:2015msa}. 
QCD evolution equations have been constructed in particular for unpolarized TMD PDFs and also for polarized TMDs such as the Sivers function. 
The numerical results obtained from the evolution equations show explicitly that QCD evolution is very significant for TMDs.  
Not only the form of the $k_\perp$-dependence, but also the width of the Gaussian evolves with $Q$.
More precisely, at small $k_\perp$, Gaussian parameterization can be used but the width evolves with $Q$. 
At larger $k_\perp$, the form of $k_\perp$-dependence is determined mainly by gluon radiation and deviates greatly from a Gaussian and also evolve with $Q$.
 In Fig.~\ref{fig:EvoGaussian}, we see an example for the evolution of the Gaussian parameterization at small $k_\perp$; 
 in Fig.~\ref{fig:EvoForm}, we see the evolution of the shape at large $k_\perp$. 
It is also important to use the comprehensive TMD evolution rather than a separate evolution of the transverse and longitudinal dependences respectively. 
We show as an example in Fig.~\ref{fig:EvoDGLAP}. 

The last thing for TMD parameterizations that we would like to mention is the TMD library (TMDlib). 
We are happy to see that, a first version has already been created~\cite{Hautmann:2014kza} in the year 2014, and updated recently.

\section{Summary and Outlook}
In summary, by comparing with what we did in studying one dimensional imaging of the nucleon with inclusive DIS, 
we presented a brief overview of our studies on three dimensional imaging of the nucleon with semi-inclusive DIS and other semi-inclusive reactions.
We summarized in particular the general form of the TMDs defined via quark-quark correlators both for TMD PDFs and FFs. 
We emphasized in particular on the theoretical framework for semi-inclusive reactions at LO pQCD but with leading and higher twist contributions consistently.  
Such theoretical framework is obtained by applying the collinear expansion technique developed in 1980s in inclusive DIS to these semi-inclusive processes. 
We summarized in particular that it applies now also to all processes where one hadron is involved. 
The results obtained in such a framework should be used as starting points for studying TMDs experimentally.

At the end, we would like to emphasize that three dimensional imaging of the nucleon is a hot and fast developing topic in last years. 
 Many progresses have been made and many questions are open. 
 We see in particular that LO pQCD leading and higher twists framework for processes 
 where one hadron is involved can be constructed using collinear expansions. 
 Factorization theorem for leading twist but with LO and higher order pQCD contributions 
 and QCD evolution equations for unpolarized TMD PDFs and the Sivers functions have also been established.  
 Especially in view of the running and planned facilities such as the electron-ion colliders, 
 we expect even rapid development in next years. 
 
 The overview is far from complete. We apologize for many aspects that we did not cover such as the generalized parton distributions, 
 the Wigner function, model calculations of TMDs, nuclear dependences, and hyperon polarization. \\
 
 \section*{Acknowledgements} 
We thank X.N. Wang, Y.K. Song, J.H. Gao and many other people for collaboration and help in preparing this review.  
ZTL thanks also John Collins and Zebo Tang for communications.
This work was supported in part by the National Natural Science Foundation of China
(Nos.11035003 and 11375104),  the Major State Basic Research Development Program in China (No. 2014CB845406) 
and the CAS Center for Excellence in Particle Physics (CCEPP).





\end{document}